\documentclass[10pt,a4paper]{amsart}
\usepackage{amsmath}
\usepackage{amscd}

\newtheorem{thm}{Theorem}[section]

\newtheorem{prop}[thm]{Proposition}

\newcommand{\thmref}[1]{Theorem~\ref{#1}}
\newcommand{\propref}[1]{Proposition~\ref{#1}}

\newsavebox{\SmallMathBox}

\def\Uu{{\mathcal U}}

\def\Ci{C^\infty}

\def\dd{\partial}

\def\Di{D\kern -.65em /}
\def\Dii{D\kern -.45em /}
\def\di{{\dd}\kern -.55em /}
\def\dii{{\dd}\kern -.40em /}

\def\noi{\noindent}

\def\Df{\mathbb{D}}
\def\Pf{\mathbb{P}}
\def\Sf{\mathbb{D}}
\def\Df{\mathbb{D}}

\def\ol{\overline}

\def\to{\rightarrow}
\def\too{\longrightarrow}

\def\wt{\widetilde}

\def\mto{\mapsto}
\def\mtoo{\longmapsto}

\def\CC{{\bf C}}

\def\MM{{\bf M}}
\def\NN{{\bf N}}

\def\Pf{\mathbb{P}}

\def\Aa{{\mathcal A}}

\def\Cc{{\mathcal C}}
\def\Dd{{\mathcal D}}
\def\Ee{{\mathcal E}}
\def\Ff{{\mathcal F}}
\def\Gg{{\mathcal G}}
\def\Hh{{\mathcal H}}
\def\Kk{{\mathcal K}}
\def\Ll{{\mathcal L}}
\def\Mm{{\mathcal M}}
\def\Nn{{\mathcal N}}

\def\Rr{{\mathcal R}}
\def\Ss{{\mathcal S}}

\def\Uu{{\mathcal U}}

\def\Ww{{\mathcal W}}
\def\Xx{{\mathcal X}}
\def\Yy{{\mathcal Y}}
\def\={\cong}
\def\>{\supset}
\def\<{\subset}
\def\ii{^{-1}}
\def\pp{^{\perp}}
\def\12{\frac{1}{2}}
\def\2{\Dd}
\def\3{\Nn}
\def\4{\Rr}
\def\6{\cup}
\def\8{\otimes}
\def\0{^{\circ}}

\def\a{\alpha}
\def\b{\beta}

\def\d{\delta}
\def\e{\epsilon}

\def\g{\gamma}

\def\k{\kappa}

\def\la{\lambda}

\def\O{\emptyset}

\def\Si{\Sigma}

\def\z{\zeta}

\def\Mor{\mbox{\rm Mor}}
\def\Ob{\mbox{\rm Ob}}

\def\Det{\mbox{\rm Det\,}}
\def\DET{\mbox{\rm DET\,}}

\def\End{\mbox{\rm End}}

\def\Hom{\mbox{\rm Hom}}

\def\Ker{\mbox{\rm Ker}}
\def\Cok{\mbox{\rm Coker}}

\def\ind{\mbox{\rm ind\,}}

\def\Ran{\mbox{\rm range}}

\def\Si{S\kern -.65em /}

\def\tr{\mbox{\rm tr\,}}
\def\Tr{\mbox{\rm Tr\,}}

\begin{document}

\title[Determinant Bundles and FQFT]{Functorial QFT, Gauge
Anomalies and the Dirac Determinant bundle}

\author{Jouko Mickelsson}

\address{Department of Theoretical Physics, Royal institute of Technology,
S-10044 Stockholm, Sweden.}

\email{jouko@theophys.kth.se}

\author{Simon Scott}

\address{Department of Mathematics, King's College,
 London WC2R 2LS, U.K.}

\email{sscott@mth.kcl.ac.uk}

\begin{abstract} Using properties of the determinant
line bundle for a family of elliptic boundary value problems, we
explain how the Fock space functor defines an axiomatic quantum
field theory which formally models the Fermionic path integral.
The `sewing axiom' of the theory arises as an algebraic pasting
law for the determinant of the Dirac operator. We show how
representations of the boundary gauge group fit into this
description and that this leads to a Fock functor description of
certain gauge anomalies.
\end{abstract}

\maketitle


\section{Introduction}

Advances in the construction of topological invariants for
low-dimensional manifolds using methods from gauge theory have led
to a great deal of interest in the construction of quantum field
theories as modified cohomology theories
\cite{At,Seg90,Seg90a,Wi1,Wi90}; that is, as generalized functors
from manifolds to vector spaces. The purpose of this paper is to
explain a construction of a functorial quantum field theory (FQFT)
using the Fock functor, generalizing a construction suggested by
Segal \cite{Seg90} in (1+1)-dimensions. This may be of particular
interest in view of recent developments in the theory of branes in
superstring theory. In doing so, we realize the higher-dimensional
gauge group representations of \cite{Mi89} in terms of a
d+1-dimensional FQFT, while the gluing law of the FQFT arises as
an algebraic pasting law for the determinant of a Dirac operator
with respect to a partition of the underlying manifold.

The aim of FQFT is to abstract the algebraic structure that the
path integral would create if it existed as a rigorous
mathematical object.  With respect to a partition of the
underlying manifold the functoriality formally encodes intriguing
formal gluing laws for spectral or topological invariants realized
as expectation values. The prototypical situation we consider is
for a family of chiral Dirac operators over an even-dimensional
manifold with closed odd-dimensional boundary. The parameter space
$\Aa$ in this case is an affine space of gauge potentials cross
Riemannian metrics, acting on which one has a group $\Gg$ of gauge
transformations. To a spin manifold $X$ with boundary $Y$ endowed
with an admissible decomposition $H_Y = W\oplus W^{\perp}$ of the
space of boundary spinor fields, the Fock functor associates a
Fock space $\Ff_W$ of {\em holomorphic} sections of a relative
determinant line bundle over the restricted Grassmannian defined
by the polarization $W$. Globally the functor associates a bundle
$\Ff$ of Fock spaces to the parameter space $\Aa$  and the sewing
properties of the determinant line bundle for a family of elliptic
boundary value problems explained in \cite{Sc98} translate into
the required functorial properties of the FQFT. The constructions
of \cite{Mi89} arise in this situation in terms of two
``orthogonal" $\Gg$-anomalies. First, there is the ``bulk"
even-dimensional chiral anomaly measuring the obstruction to a
$\Gg$-equivariant determinant regularization. Second, associated
to the gauge group on the boundary one has the odd-dimensional
Mickelsson-Faddeev commutator anomaly. In the FQFT context the
$\Gg$-action is lifted from $\Aa$ to a projective {\em bundle map}
on the Fock bundle, rather than an automorphism of the whole
(fixed) space of sections of the determinant line bundle
\cite{Mi89}.  FQFT formally encodes the relation between the path
integral and Hamiltonian approaches to second quantization, the
Fock functor FQFT we consider provides a coherent framework in
which to describe simultaneously the path integral (determinant
line) description of gauge anomalies and their Hamiltonian (Fock
space) realization. We hope this may serve to clarify some of the
underlying mathematical structures of the QFT.

In the remainder of the Introduction we recall from
\cite{At,Seg90a,Wi90} the axiomatic characterization of a QFT and
the heuristic path integral formulae this aims to encode. In
Section 2 we explain some facts about determinant bundles and Fock
spaces associated to families of elliptic boundary value problems.
In Section 3 we define the Fock space functor in general and
outline its fundamental properties. In Section 4 we apply this to
the interactive Yang-Mills gauge theory associated to $\Aa$ and
discuss the boundary gauge group action and the chiral anomaly and
commutator anomalies. In Section 5 we outline the path integral
formulae for an elliptic boundary value problem which the Fock
functor aims to model, and present a concrete $0+1$-dimensional
example which relates our constructions to the finite-dimensional
Fermionic (Berezin) integral.

\subsection{Axiomatic QFT}\label{subsec:aqft}

A (d+1)-dimensional FQFT means a functor from
 the category $C_{d}$ of $d$-dimensional closed manifolds and cobordisms
 to the category of vector spaces and linear maps, which
satisfies certain natural axioms suggested by path integral
formulae. A morphism in $C_{d}$ between $d$-dimensional manifolds
$Y_{0},Y_{1}$ is a $(d+1)$-dimensional manifold $X$ with boundary
$\dd X = Y_{0}\sqcup Y_{1}$.  The orientation on the `incoming'
boundary $Y_0$ is assumed to be induced by the orientation of $X$
and the inward directed normal vector field on the boundary,
whereas for the `outgoing' boundary $Y_1$ the orientation is fixed
by the outward directed normal vector field. Let $C_{vect}$ denote
the category whose objects are topological vector spaces and whose
morphisms are homomorphisms.

A $d+1$-dimensional FQFT means a functor $Z:C_{d}\to C_{vect}$
assigning to each $d$-dimensional manifold $Y$ a vector space
$Z(Y)$ and to each cobordism $X$ a vector $Z_{X}\in Z(\dd X)$. By
{\it fiat} $ Z(\emptyset) = \mathbb{C}$, so that if $X$ is closed
then $Z_{X}$ is a complex number. $Z$ is required to satisfy the
following axioms.

\bigskip

\noi For  $(d+1)$-dimensional manifolds $X,X_{0},X_{1}$ and
$d$-dimensional manifolds $Y,Y_{0},Y_{1}$:

\begin{description}
  \item[A1.\;Multiplicativity]
$\;\;Z_{X_0\sqcup X_1} = Z_{X_0}\otimes Z_{X_1},\;\;\;\;
Z(Y_0\sqcup Y_1) = Z(Y_0)\otimes Z(Y_1).$
  \item[A2.\;Duality]If $\ol{Y}$ denotes
  $Y$ with reversed orientation then
  $\;\;Z(\ol{Y}) = Z(Y)^{*}.$
  \item[A3.\;Associativity] If $M=X_{0}\cup_{Y}X_{1}$ with
  $\dd X_{0} = \ol{Y_0}\sqcup Y$ and $\dd X_{1} = \ol{Y}\sqcup
  Y_{1}$, then  $$ Z_M = Z_{X_{1}}\circ Z_{X_{0}}.$$
  \item[A4.\;Hermitian] $\;\;Z_{\ol{X}} = \ol{Z_X}.$
\end{description}

\noi The associativity property refers to the fact that axioms A1
and A2 mean that a cobordism $X\in C_d$ induces a linear
transformation $Z_X\in C_{vect}$ through the identifications
$$Z_X\in  Z(\ol{Y_0}\sqcup Y_1) = Z(\ol{Y_0})\otimes Z(Y_1) =
Z(Y_0)^{*}\otimes Z(Y_1) = \Hom (Z(Y_0),Z(Y_1)).$$ \noi Thus
morphisms in $C_d$ are taken to morphisms in $C_{vect}$. In
particular, we have then a canonical pairing $Z(\ol{Y_0})\otimes
Z(Y)\otimes Z(\ol{Y})\otimes Z(Y_1) \too Z(\ol{Y_0})\otimes
Z(Y_1).$ In the case when $Y_0 = Y_1 = \emptyset$, so $M$ is a
closed manifold , this becomes a pairing
\begin{equation}\label{e:pairing1}
(\, , \,) : Z(Y)\otimes Z(\ol{Y})\too \mathbb{C},
\end{equation}
\noi and A3 implies the {\em sewing property}
\begin{equation}\label{e:sewing}
  Z_M = (Z_{X_{0}},Z_{X_{1}}).
\end{equation}

This is perhaps the most striking feature of a FQFT, it states
that by partitioning the manifold $M$ into `simpler' codimension 0
submanifolds, the number $Z_M$ can be computed by evaluating over
the submanifolds and then sewing together the results via the
bilinear pairing. The bilinear pairing further implies that if
$\dd X = Y\sqcup \ol{Y}$ and $f:Y\to Y$ is an orientation
reversing diffeomorphism, then
\begin{equation}\label{e:trace}
\Tr (Z_{X}(f)) = Z_{X_{f}}.
\end{equation}
\noi Here $X_{f}$ is the closed manifold obtained by identifying
the boundary components via $f$, and $Z_{X}(f)\in\End (Z(Y))$ is
induced by functoriality (and in \eqref{e:trace} is implicitly
assumed to be trace class).

The Hermitian axiom A4 applies to the case of a unitary FQFT. this
means there is a non-degenerate Hermitian structure $<\, , \, >
:Z(Y)\otimes \ol{Z(Y)} \too \mathbb{C},$ and hence a canonical
isomorphism $Z(Y)^{*}\equiv \ol{Z(Y)}$. A4 is the corresponding
expected behaviour of $Z_X$.

\bigskip
These axioms are `idealized', and in practice some modifications
are needed. This is illustrated in the FQFT we consider in Section
3.

\subsection{Heuristic Path Integral Formulae}

The above framework aims to algebraicize the relation between the
Feynman path integral formulation of QFT and its Hilbert space
formulation. The following heuristic interpretation  is useful to
bear in mind. If $X$ has connected boundary, the vector $Z_X$
represents the partition function, which is given by a formal path
integral
\begin{equation}\label{e:pi}
Z_X : \Ee (Y) \too \mathbb{C},\;\;\;Z_X (f) =
\int_{\Ee_{f}(X)}e^{-S(\psi)}\Dd\psi,
\end{equation}
where $\Dd\psi$ is a formal measure. Here
$S:\Ee_{f}(X)\to\mathbb{C}$ is an action functional on a space of
fields on $X$, which for definiteness we shall take to be the
space of $C^{0}$ functions on $X$, with boundary value $f\in
\Ee(Y)$. The vector space $Z(Y)$ is a space of functions on $\Ee
(Y)$ and forms the Hilbert space of the theory, and $Z_X$ is the
vacuum state.

To a cobordism $X\in C_d$ with $\dd X = \ol{Y_0}\sqcup Y_0$ one
has $f=(f_0,f_1)$, and then

$$ Z_X(f_0,f_1) : \int_{\Ee_{(f_0,f_1)}(X)}e^{-S(\psi)}\Dd\psi,$$

\noi is the kernel of the linear operator $ Z_X \in \Hom
(Z(Y_0),Z(Y_1))$ defined by

$$ Z_X(\xi_0)(f_1) = \int_{\Ee (Y_0)} Z_X(f_0,f_1)\xi_0(f_0)\Dd
f_0.$$

If $Y_0 = Y_1 = Y$ we hence obtain the bilinear form on
$Z(\ol{Y})\times Z(Y)$ corresponding to \eqref{e:pairing1}:

$$<\xi_0,\xi_1> = \int_{\Ee (Y)}\xi_{1}(f)Z_{X}(\xi_0)(f)\,\Dd
f.$$

\noi In the case of a closed manifold $M=X_{0}\cup_{Y}X_{1}$ we
can express the space of $C^{0}$ functions on $M$ as a fibre
product $\Ee (M) = \Ee (X_0)\times_{\Ee (Y)}\Ee (X_1)$ and so
formally one expects an equality

\begin{eqnarray}\label{e:pisewing2}
\int_{\Ee (M)}e^{-S(\psi)}\Dd\psi & = & \int_{\Ee (Y)}\Dd f
\int_{\Ee_{f}(X_0)}e^{-S(\psi_0)}\Dd\psi_0\int_{\Ee_{f}(X_1)}e^{-S(\psi_1)}\Dd\psi_1
\nonumber\\ & = &  \int_{\Ee (Y)}Z_{X_{0}} (f)Z_{X_{1}} (f)\Dd f.
\end{eqnarray}

\noi which is the path integral version of the algebraic sewing
formula \eqref{e:sewing}. The Hamiltonian of the theory is defined
by the Euclidean time evolution operator $e^{-tH} = Z_{Y\times
[0,t]}\in \End (Z(Y)),$ and to compute the trace one has the
integral formulae $$\Tr (e^{-tH}) = \int_{\Ee (Y)}\Tr Z_{X}(f,f)\,
\Dd f,$$ and corresponding to \eqref{e:pi}
\begin{equation}\label{eTrH2}
\Tr (e^{-tH(f)}) = \int_{\Ee (X_f)}e^{-S(\psi)}\, \Dd\psi.
\end{equation}

\bigskip

The sewing formula \eqref{e:pisewing} says that the partition
function on $M$ is the vacuum-vacuum expectation value calculated
from the partition functions on the two halves. Equivalently: the
invariant $Z_M$ is obtained from $Z_{X_{0}}(f)$ and $Z_{X_{1}}
(f)$ by integrating (`averaging') away the choice of boundary data
$f$. In the case of determinants of Dirac operators this formalism
provides some insightful to sewing formulae  relative to a
partition of the underlying manifold (see Section 5). First, we
need to review some facts about determinant and Fock bundles for
families of Dirac operators.


\section{Determinant line bundles and Fock spaces}

The determinant of a family of first-order elliptic operators
arises canonically not as a function, but as a section of a
complex line bundle called the determinant line bundle. The
anomalies we shall discuss may be realized as obstructions to
constructing appropriate trivializations of that bundle.
Equivalently, we can view the determinant line of an operator as a
ray in the associated Fock space (via the `Pl\"ucker embedding'),
and globally the determinant bundle as rank 1 subbundle of an
infinite-dimensional Fock bundle to which the gauge group lifts as
a projective bundle map.

First, recall the construction of the determinant line bundle for
a family of Dirac-type operators over a closed compact manifold
$M$. Such a family can be specified by a smooth fibration of
manifolds $\pi : M\too B$ with fibre diffeomorphic to $M$, endowed
with a Riemannian metric $g_{M/B}$ along the fibres and a vertical
bundle of Clifford modules $S(M/B)$ which we may identify with the
vertical spinor bundle tensored with an external vertical gauge
bundle $\xi$. We assume that $\xi$ is endowed with a Hermitian
structure with compatible connection. The manifold $B$ is not
required to be compact. We refer to this data as a {\em geometric
fibration}.

Associated to a geometric fibration one has a smooth elliptic
family of Dirac operators $\Df=\{ D_{b}:b\in B \}:\Hh\too\Hh$,
where $\Hh = \pi_{*}(\Ss(M/B))$ is the infinite-dimensional
Hermitian vector bundle on $B$ whose fibre at $b$ is the Frechet
space of smooth sections $\Hh_{b} = C^{\infty}(M_{b},\Ss_{b})$,
where $\Ss_b$ is the appropriate Clifford bundle. If $M$ is
even-dimensional there is a $\mathbb{Z}_{2}$ bundle grading $\Hh=
\Ff^{+}\oplus\Ff^{-}$ into positive and negative chirality fields
and we then have a family of chiral Dirac operators $\Df:
\Ff^{+}\too\Ff^{-}$. The Quillen determinant line bundle
$\DET(\Df)$ is a complex line bundle over $B$ with fibre at $b$
canonically isomorphic to the complex line $\Det(\Ker D_{b})^{*}
\otimes \Det\Cok(D_{b})$ \cite{BiFr86,Qu85}, where for a
finite-dimensional vector space $V$, $\Det V$ is the complex line
$\wedge^{max}V$. The bundle structure is defined relative to the
covering of $B$ by open subsets $U_{\la}$, with $\la\in R^{+}$,
parameterising those operators $D_{b}$ for which $\la$ is not in
the spectrum of the Laplacian $D_{b}^{*}D_{b}$. Over each
$U_{\la}$ are smooth finite-rank vector bundles
$H^{+}_{\la},H^{-}_{\la}$ equal to the sum of eigenspaces of
$D_{b}^{*}D_{b}$ (resp. of $D_{b}D_{b}^{*}$) for eigenvalues less
than $\la$, and one defines
\begin{equation}
\DET(\Df)_{|U_{\la}} = \Det (H^{+}_{\la})^{*}\otimes \Det
H^{-}_{\la}.
\end{equation}
The locally defined line bundles patch together over the overlaps
$U_{\la}\cap U_{\la^{'}}$ in a natural way. This `spectral'
construction of the determinant line bundle is designed to allow
one to define the Quillen $\z$-function metric
 and a compatible connection whose
curvature $R^{\z}$ is identified with the 2-form component of the
Bismut family's index density:
\begin{equation}\label{e:zcurv}
R^{\z} = (2\pi
i)^{-n/2}\left[\int_{M/B}\widehat{A}(M/B)ch(\xi)\right]_{2},
\end{equation}
where $\widehat{A}(M/B)$ is the vertical A-hat form and $ch$ the
Chern character, see \cite{Qu85,BiFr86,BeGeVe}.

There is, however, a natural alternative construction of the
determinant line bundle, due to Segal \cite{SW85, PrSe86}, and
applied to Dirac families in \cite{Sc98}, which allows us to
consider more general smooth families of Fredholm operators, which
need not be elliptic operators. Let $\a:H^0\to H^1$ be a Fredholm
operator of index zero. Then a point of the complex line  over
$\a$ is an equivalence class $[A,\lambda ]$ of pairs $(A,\lambda
)$, where the operator $A:H^0\to H^1$ is such that $A- \a \in
\End(H^0)$ is trace-class, $\lambda\in\mathbb{C},$ and the
equivalence relation is defined by $(Aq,\lambda ) \sim
(A,\det_{F}(q)\lambda )$ for $q\in \End(H^{0})$ an operator of the
form identity plus trace-class, and $\det_{F}$ denotes the
Fredholm determinant. If $\ind \a = d$ we define $\Det (\a) :=
\Det(\a\oplus 0)$  with $\a\oplus 0$ acting $H^0\to H^{1} \oplus
\mathbb{C}^{d}$ if  $d
> 0$, or $H^0 \oplus \mathbb{C}^{-d}\too H^{1}$ if  $d < 0$.  Note
that, by definition, a Fredholm operator of index zero has an
approximation by an invertible operator $A$ such that $A-\alpha$
has finite rank. We work with the larger ideal of trace-class
operators in order to be able to use the (complete) topology
determined by the trace norm.

The abstract determinant of $\a$ is defined to be the canonical
element $\det \a := [\a,1]\in \Det (\a)$. For an admissible smooth
family of Fredholm operators $\Aa = \{\a_{b}:b\in B\}:\Hh^{0}\to
\Hh^1$ acting between (weak) vector bundles $\Hh^i$
\cite{Sc98,SW85}, the union $\DET(\Aa)$ of the determinant lines
is naturally a complex line bundle. The bundle structure is
defined relative to a denumerable open covering of open sets
$U_{\tau}$, where $\tau:\Hh^{0}\to \Hh^1$ is finite-rank and
$U_{\tau}$ parameterizes those $b$ for which $\a_{b} + \tau_b$ is
invertible, via the local trivialization $b\mapsto det(\a_{b} +
\tau_b)$ over $U_{\tau}$. On the intersection $U_{\tau^1}\cap
U_{\tau^2}$ the transition function is $b\to det_{F}((\a_{b} +
\tau^{1}_{b})\ii (\a_{b} + \tau^{2}_{b}))$. For a family of
elliptic operators, such as $\Df$, there is a canonical
isomorphism between the two constructions of the determinant
bundle described above which preserves the determinant section
$b\mapsto\det D_{b}$, and we may therefore use them
interchangeably \cite{Sc98}.

This is important when we consider the determinant line bundle for
a family of elliptic boundary value problems (EBVPs). To define
such a family we proceed initially as for the case of a closed
manifold with a geometric fibration $\pi : M\too B$ of connected
manifolds, but with fibre diffeomorphic to a compact connected
manifold $X$ with boundary $\dd X = Y$. Note that the boundary
manifolds $\dd M$ and $\dd X$ may possibly be disconnected.
Globally we obtain as before a family of Dirac operators $\Df=\{
D_{b}:b\in B \}:\Hh\too\Hh$. We assume that the geometry in a
neighbourhood $\Uu\equiv \dd M \times [0,1]$ of the boundary is a
pull-back of the geometry induced on the boundary geometric
fibration of closed boundary manifolds $\dd\pi : \dd M\too B$.
This means that all metrics and connections on $T(M/B)$ and
$S(M/B)$ restricted to $U$ are geometric products composed of the
trivial geometry in the normal $u$-coordinate direction, and the
boundary geometry in tangential directions, so $g_{M/B} = du^2 +
g_{\dd M/B}$ and so forth. In $\Uu_b := U\equiv \dd X_b \times
[0,1] $ the Dirac operator $D_{b}$ then has the form
\begin{equation}\label{e:Du}
  D_{b|U} = G_b\left(\frac{\dd}{\dd u} + D_{Y,b}\right),
\end{equation}
\noi  where $D_{Y,b}$ is a boundary Dirac operator and $G_b$ is a
unitary bundle automorphism, the Clifford multiplication related
to the outward directed vector field $\frac{\dd}{\dd u}.$ The
family of boundary Dirac operators $\Df_{Y} = \{ D_{Y,b}:b\in B
\}:\Hh_Y\too\Hh_Y$, where $\Hh_Y$ is the bundle with fibre $\Ci
(Y_b,S_{Y_b})$ at $b\in B$, is identified with family defined by
the fibration $\dd\pi : \dd M\too B$.

In contrast to the closed manifold case, the operators $D_b$ are
not Fredholm. The crucial analytical property underlying the
following determinant line and Fock bundle identifications is the
existence of a canonical identification between the
infinite-dimensional space $\Ker(D_{b})$ of solutions to the Dirac
operator and the boundary traces $K(D_b) = \gamma\Ker(D_{b})$,
where $\gamma : \Ci (X_b,S_b)\to \Ci (Y_b,S_{Y_b})$ is the
operator restricting sections to the boundary. More precisely, the
{\it Poisson operator} $\Kk_b :
 \Ci (Y_b,S_{Y_b})\to \Ci(X_b,S_b)$ restricts to define the above
 isomorphism. It extends to a continuous operator
$\Kk_b : H^{s-1/2}(Y_b;S|Y_b) \to H^s(X_b;S)$ on the Sobolev
completions with range $$ker(D_b,s) = \{f \in H^s(X_b;S) \ : \ D_b
f = 0 \ in \ X_b\setminus Y_b \} \ \ ,$$ and $\Kk^{s}_{b} :
K(D_b,s)) \to ker(D_b,s)$ is an isomorphism (see \cite{BoWo93,
Gr99}). The {\em Poisson operator} of  $D_b$ defines the {\em
Calderon projection}:
\begin{equation}\label{e:calderon}
P(D_b) = \g\Kk^{1}_b.
\end{equation}
$P(D_b)$ is a pseudodifferential projection on
$L^{2}(Y_{b},S_{Y_{b}})$ which we can take to be orthogonal with
range equal to $ker(D,s)$. The construction depends smoothly on
the parameter $b\in B$ and so globally we obtain a smooth map
$P(\Df):B\to \End(\Hh_{Y})$ defining, equivalently, a smooth
Frechet subbundle $K(\Df)$ of $\Hh_Y$ with fibre $K(D_b)=
\Ran(P(D_b))$. Because of the tubular boundary geometry, $P(D_b)$
in fact differs from the APS spectral projection $\Pi_b$ by only a
smoothing operator, see \cite{Sc95,Gr99}. Recall that there is a
polarization $H_{Y_{b}} = H_{b}^{+}\oplus H_{b}^{+}$ into the
non-negative and negative energy modes of the elliptic
self-adjoint boundary Dirac operator $D_{Y,b}$ and $\Pi_b$ is
defined to be the orthogonal projection onto $H_{b}^{+}$. Hence
$P(D_b)$ is certainly an element of the Hilbert-Schmidt
Grassmannian $Gr_{b}$ parameterizing projections on $H_{Y,b}$
which differ from $\Pi_{b}$ by a Hilbert-Schmidt operator, where
by projection we mean self-adjoint indempotent. Associated to
$P\in Gr_{b}$ we have the elliptic boundary value problem (EBVP)
for $D_{b}$
\begin{equation}\label{e:ebvp}
  D_{P,b} = D_{b} : dom(D_{P,b}) \to L^2(X_{b};S^{1})
\end{equation}
\noi with domain $ dom \ D_{P,b} = \{s \in H^1(X_b;S_{b}^{0}) \ :
\ P(s|Y_b) = 0 \} $. The operator $D_{P,b}$ is Fredholm with
kernel and cokernel consisting of smooth sections, see
\cite{BoWo93} for a general account of EBVPs in index theory. The
smooth family of EBVPs $D_{Gr_{b}}:= \{D_{P,b}:P\in Gr_b\}$
defines an admissible family of Fredholm operators and hence an
associated determinant line bundle $\DET(D_{Gr_{b}})\to Gr_{b}$.
On the other hand,  for each choice of a basepoint $P_0\in Gr_b$
we have the smooth family of Fredholm operators $$\{ P_{W_{0}, W}
:= P\circ P_{0}:W_0\to W\;:P\in Gr_b\},$$ where $ran(P_0)=W_0,
ran(P)= W$,and hence a relative (Segal) determinant line bundle
$\DET_{W_{0}}\to Gr_{b}$ based at $W_0$. The bundles so defined
for different choices of basepoint are all isomorphic, but not
quite canonically. More precisely, from \cite{Seg90,Sc98}, given
$P_0,P_1\in Gr_b$ there is a canonical line bundle isomorphism
\begin{equation}\label{e:isom1}
  \DET_{W_{0}} \cong \DET_{W_{1}}\otimes \DET(W_{0},W_{1}),
\end{equation}
where $\DET(W_{0},W_{1})$ means the trivial line bundle with fibre
the relative determinant line $\DET(W_{0},W_{1}) :=
\Det(P_{W_{0},W_{1}}) $. In view of the identification defined by
the Poisson operator it is perhaps not surprising that the
determinant line bundle $\DET(D_{Gr_{b}})$ is classified by the
basepoint $K_b$:
 there is a canonical line bundle isomorphism
\begin{equation}\label{e:isom2}
 \DET(D_{Gr_{b}}) \cong \DET_{K_b},
\end{equation}
preserving the determinant sections $$\det(D_{Gr_{b}})
\longleftrightarrow \det (S_b (P)),$$ where $S_b (P) :=
PP(D_b):K(D_b) \to ran(P).$ (see \cite{Sc98}). To translate these
facts into global statements for operators parameterized by $B$ we
require the notion of a {\em spectral section}, or {\em Grassmann
section} \cite{Sc98} (we may use both names, the latter is
sometimes more appropriate in more general situations). For each
$b\in B$ we have the restricted Grassmannian $Gr_b$ and globally
these Hilbert manifolds fit together to define a fibration $\Gg
r_{Y}\to B$. A spectral section $\Pf = \{P_b:b\in B\}$ for the
family $\Df$ is defined to be a smooth section of that fibration,
and we denote the space of sections by $Gr(M/B)$. By cobordism,
such sections always exist. In particular, the family of Dirac
operators $\Df$ defines canonically the Calderon section
$P(\Df)\in Gr(M/B)$. In this sense one may think of the parameter
space $B$ as a `generalized Grassmannian' (i.e. parameterizing the
subspaces $K(D_b)$), and the usual Grassmannian as a `universal
moduli space'. Notice, however, that the map $b\to \Pi_b$ is
generically not a smooth spectral section because of the flow of
eigenvalues of the boundary family. Indeed, it is this elementary
fact that is the source of gauge anomalies, see \cite{Mi90} and
Section 4. A spectral section has a number of consequences for
determinants:\\[2mm] {\bf First}: We obtain a smooth family of
EBVPs $(\Df, \Pf) = \{D_{P,b} : = (D_b)_{P_b} :b\in B\}$ which has
an associated determinant line bundle $\DET(\Df,\Pf)\to B$ with
determinant section $b\mapsto D_{P,b}$.\\[1mm] {\bf Second}: A
spectral section $\Pf$ defines a smooth infinite-dimensional
vector bundle $\Ww$ with fibre $W_b = \Ran(P_b)$, and associated
to $\Pf$ we have the smooth family of Fredholm operators
$\Sf(\Pf):K(\Df)\to \Ww$, parameterizing the operators $S(P_b) :=
P_{K(D_b),W_b}:K(D_b)\to W_b$. This also has a determinant line
bundle $\DET(\Sf(\Pf))$, and corresponding to \eqref{e:isom2},
there is a canonical line bundle isomorphism
\begin{equation}\label{e:isom3}
 \DET(\Df,\Pf) \cong \DET(\Sf(\Pf)),
  \hskip 5mm det(D_{P,b})\leftrightarrow det(S(P_{b})),
\end{equation}
preserving the determinant sections. Given a pair of sections
$\Pf_1,\Pf_2\in Gr(M/B)$ there is the smooth family of admissible
Fredholm operators $(\Pf^1,\Pf^2):\Ww^1\to\Ww^2$, and
corresponding to \eqref{e:isom1} and \eqref{e:isom3} one finds a
canonical isomorphism
\begin{equation}\label{e:isom4}
  \DET(\Df,\Pf^1) \cong \DET(\Df,\Pf^2)\otimes \DET(\Pf^1,\Pf^2),
\end{equation}
(which does not preserve the determinant sections). We refer to
\cite{Sc98} for details.
\\[1mm] {\bf Third}: We obtain a bundle of Fock spaces $\Ff_{\Pf}$
over $B$. To see this, return for a moment to the case of a single
operator and its Grassmannian $Gr_b$. By choosing a basepoint
$P_0\in Gr_b$, we obtain the determinant line bundle
$\DET_{W_{0}}\to Gr_{b}$. This is a holomorphic line bundle, but
has no global holomorphic sections. The dual  bundle
$\DET_{W_{0}}^{*}$, on the other hand, has an infinite-dimensional
space of holomorphic sections, and this, by definition, is the
Fock space based at $W_0$:
\begin{equation}\label{e:fock1}
  \Ff_{W_0,b} := \Gamma_{hol}(Gr_{b};\DET_{W_{0}}^{*}).
\end{equation}
Actually, a Fock space comes together with a vacuum vector and a
representation of the canonical anticommutation relations; we
shall return to this at the end of the section. Taking the union
$\Ff_{b} := \cup_{W\in Gr_b}\Ff_{W,b}$ we obtain the Fock bundle
over $Gr_b$. This bundle is topologically completely determined by
`the' determinant bundle $\DET_{W_{0}}$, in fact this is the most
direct way to define the bundle structure on $\Ff_b$. To be
precise, if we change the basepoint we find, dropping the $b$
subscript, a canonical isomorphism
\begin{eqnarray*}
  \Ff_{W_1} & = & \Gamma_{hol}(Gr;\DET_{W_{1}}^{*}) \\
  & \cong &  \Gamma_{hol}(Gr_{b};\DET_{W_{0}}^{*}\otimes
  \DET(W_{1},W_{0})^*)\\
& \cong &  \Gamma_{hol}(Gr_{b};\DET_{W_{0}}^{*})\otimes
  \DET(W_{1},W_{0})^*\\
& \cong &  \Ff_{W_0}\otimes
  \DET(W_{0},W_{1}),\label{e:fock2}
\end{eqnarray*}
where we use \eqref{e:isom1}. Hence relative to a basepoint
$W_0\in Gr_b$ we have a canonical isomorphism
\begin{equation}\label{e:Fdet}
  \Ff_b \cong \Ff_{W_0}\otimes\DET_{W_{0}},
\end{equation}
where the first factor on the right-side is the trivial bundle
with fibre $\Ff_{W_0}$. Hence the topological type of the Fock
bundle $\Ff_b$ is determined by that of the determinant line
bundle $\DET_{W_{0}}$ for any basepoint $W_0$. One moves between
the isomorphisms for different basepoints via \eqref{e:isom1}.

As an abstract vector bundle, a Fock bundle is always trivial (but
not necessarily canonically); this is because of the fact that
(according to Kuiper's theorem) the unitary group in an
infinite-dimensional Hilbert space is contractible. However, as
already mentioned, the Fock spaces are equipped with additional
structure, the vacuum vectors related to a choice of a family of
(Dirac) Hamiltonians, which will modify this statement. In the
case of \eqref{e:Fdet} we have a preferred line bundle (the
`vacuum bundle') inside of the Hilbert bundle and the structure
group is reduced giving a nontrivial Fock bundle; this will be
discussed in more detail in section 4.

Now as we let $b$ vary we obtain a vertical Fock bundle $\Ff(M/B)$
over the total space $\Gg r_Y$ of the Grassmann fibration, which
restricted to the fibre $Gr_b$ of $\Gg r_Y$ coincides with
$\Ff_{b}$. The bundle structure is a obvious from the local
triviality of the fibration $\Gg r_Y\to B$. A spectral section
$\Pf$ is a smooth cross section of that fibration, and hence by
pull-back we get a Fock bundle  over $B$ associated to $\Pf$:
\begin{equation}\label{e:fock4}
  \Ff_{\Pf} := \Pf^{*}(\Ff(M/B)) \too B,
\end{equation}
with  fibre   $\Ff_{P_b} =
\Gamma_{hol}(Gr_b;\DET_{W_{b}}^{*}),\;\; W_b =
  ran(P_b)$  at $b\in B$.
In the following we may at  times also write $\Ff_{\Pf} =
\Ff_{\Ww},$ where $\Ww\to B$ is the bundle associated to $\Pf$.
Moreover, from the equivalences above we get that the various Fock
bundles are related in the following way.
\begin{prop}\label{p:fock1}
For spectral sections $\Pf^1,\Pf^2\in\Gg r(M/B)$, there is a
canonical isomorphism of Fock bundles
\begin{equation}\label{e:fock6}
  \Ff_{\Pf^1} \cong \Ff_{\Pf^2}\otimes \DET(\Pf^1,\Pf^2).
\end{equation}
\end{prop}

Notice, in a similar way to $\Ff_{\Pf}$,  we can also identify the
determinant line bundle $\DET(\Pf^1,\Pf^2)$ as a pull-back bundle.
For associated to the section $\Pf^1$ we have a vertical
determinant line bundle $\DET_{\Pf^1}\to\Gg r_Y$, which restricts
to $\DET_{W^{1}_{b}}$ over $Gr_b$, where $W^{1}_{b} =
ran(P^{1}_{b})$. Then $\DET(\Pf^1,\Pf^2)=
(\Pf^{2})^{*}(\DET_{\Pf^1})$. In particular, associated to the
family $\Df$ of Dirac operators parameterized by $B$, we have the
canonical spectral section $P(\Df)$, and by \eqref{e:isom3} we
have a canonical isomorphism $\DET(\Df,\Pf)\cong
\Pf^{*}(\DET_{P(\Df)})$. At the Fock space level we have a Fock
bundle $\Ff_{\Df}$ canonically associated to the family $\Df$,
independently of an extrinsic choice of spectral section, whose
fibre at $b\in B$ is $\Ff_{D_b} := \Gamma_{hol}(Gr_{b};
\DET(D_{Gr_{b}})^*)$. From \eqref{e:isom2} and \eqref{e:fock6} we
obtain the Fock space version of \eqref{e:isom3} and
\eqref{e:isom4}:
\begin{prop}\label{p:fock2}
There is a canonical isomorphism of Fock bundles
\begin{equation}\label{e:fock7}
  \Ff_{\Df} \cong \Ff_{P(\Df)}.
\end{equation}
For $\Pf\in \Gg r(M/B)$:
\begin{equation}\label{e:fock8}
  \Ff_{\Df} \cong \Ff_{\Pf}\otimes \DET(\Df,\Pf).
\end{equation}
\end{prop}

\noi Thus the topology of Fock bundle and the  determinant line
bundle are intimately related. This is the topological reason
relating the Schwinger terms in the Hamiltonian anomaly to the
index density.

The Fock space $\Ff_{W_0}$ based at $W_0\in Gr_b$ can be thought
of more concretely in terms of equivariant functions on the
Stiefel frame bundle  over $Gr_b$. To describe this we fix an
orthonormal basis of eigenvectors of $D_{Y,b}$ in $H_{Y_{b}}$ such
that $e_i\in H_{b}^{-}$ for $i\leq 0$ and $e_i \in H_{b}^{+}$ for
$i>0$.  A point in the fibre  over $W\in Gr_b$ of the Stiefel
bundle $St_b$ based at $H_{b}^{+}$ is a linear isomorphism $\xi :
H_{b}^{+} \to W$ such that $\Pi_b\circ \xi : H_{b}^{+}\to
H_{b}^{+}$ has a Fredholm determinant. $\xi$ is also referred to
as an `admissible basis' for $W$ (relative to $H_{b}^{+}$), in so
far as it transforms $e_i, i>0$ to a basis for $W$. $\xi$ can be
thought of as a matrix $\begin{pmatrix}
    \xi_{+} \\
     \xi_{-} \
  \end{pmatrix}$
with columns labeling the elements of the basis and rows the
coordinates in the standard basis $e_i$. If $\xi, \xi^{'}$ are two
admissible bases for $W$ then $\xi^{'} = \xi.g$ where $g$ is an
element of the restricted general linear group $Gl^1$ consisting
of invertible linear maps $g:{H_b}^+ \to {H_b}^+$ such that $g-I$
is trace-class. $Gl^1$ acts freely on $St_b\times \mathbb{C}$ by
$(\xi,\la).g = (\xi.g,\la det_{F}(g)^{-1})$ and we obtain the
alternative construction of
\begin{equation}\label{e:Stdet}
  \DET_{H_{b}^{+}} = St_b\times_{Gl^1}\mathbb{C}.
\end{equation}
Similarly, for $W\in Gr_b$ we have $ \DET_{W} =
St_W\times_{Gl^1}\mathbb{C}$, where $St_W$ is the corresponding
frame bundle based at $W$. In particular, notice that an an
isomorphism $St_{W_0}\to St_{W_1}$ is specified by an invertible
operator $A:W_0\to W_1$ such that $P_1 P_0 - A$ is trace-class,
from which we once again have the identification \eqref{e:isom1}.

In this description, an element of the Fock space $\Ff_{W_b}$ is a
holomorphic function $\psi : St_{W_{b}}\to \mathbb{C}$
transforming equivariantly under the $Gl^1$ action as $\psi(\xi.g)
= \psi(\xi) det_{F}(g)$. The distinguished element
\begin{equation}\label{e:vac1}
\nu_{W_b}(\xi) = det_{F}(P_{W_b}\xi)
\end{equation}
is the {\em vacuum vector}. Equivalently, an element of
$\Ff_{W_b}$ is a holomorphic function $f : \DET_{W_{b}}\to
\mathbb{C}$ which is linear on each fibre, and from this view
point the vacuum vector is the function
\begin{equation}\label{e:vac2}
\nu_{W_b}([\a,\la]) = \la det_{F}(P_{W_b}\a),
\end{equation}
for any representative $(\a,\la)$ of the equivalence class
$[\a,\la]\in \Det(W_b,W)$.

A generalization of this leads to the Pl\"ucker embedding. First,
for $W\in Gr_b$, fix an orthonormal basis $\{e_i\}_{i\in\Bbb Z}$
of $H_b$ such that $e_i\in W^{\perp}$ for $i\leq 0$ and $e_i \in
W$ for $i>0;$. Let $\Ss$ be the set of all increasing sequences of
integers $S =(i_1,i_2,\dots)$ with $S-\mathbb{N}$ and
$\mathbb{N}-S$ finite. For each sequence $S$ we have an admissible
basis $\xi(S) = \{e_{i_1}, e_{i_2},\dots \}\in St_{W}$, and the
Fredholm index of the operator $P_{S}P_W:W \to H_{S}$, where
$H_{S}$ is the closed subspace spanned by $\xi(S)$ and $P_{S}$ the
corresponding orthogonal projection, defines a bijection
$\pi_{0}(Gr_b)\to \mathbb{Z}$. The Pl\"ucker coordinates of the
basis $\omega\in St_W$ are the collection of complex numbers
$\psi_{S}(\omega) = det_{F}(P_S \omega) = det_{F}(\omega_{S})$,
where $\omega_{S}$ is the matrix formed from the rows of $\omega =
\begin{pmatrix}
    \omega_{+} \\
     \omega_{-} \
  \end{pmatrix}$ labeled by $S$. In particular
  $\psi_{ \mathbb{N}}(\omega)$ is the coordinate
defined by the vacuum vector. If $\omega$ is a basis for $W^{'}\in
Gr_{b}$, then the Pl\"ucker coordinates of a second admissible
basis $\omega_1$ differ from those of $\omega$ by the Fredholm
determinant of the matrix relating the two bases. The Pl\"ucker
coordinates therefore define a projective embedding $Gr_b\to
\Ff_W$. This is prescribed equivalently by the map
\begin{equation}\label{e:phi1}
  \phi : St_{W_b} \times St_{W_b} \too \mathbb{C},\hskip 5mm
  \phi(\tau,\omega) = det_{F}(\tau^{*}\omega)
   = det_{F}(\tau_{+}^{*}\omega_{+} + \tau_{-}^{*}\omega_{-}),
\end{equation}
with respect to which the Pl\"ucker coordinates are
 $\psi_{S}(\omega) = \phi(\xi(S),\omega)$. $\phi$
is the same thing as the map on the determinant bundle
\begin{equation}\label{e:phi2}
  g_{\phi} : \DET_{W_b} \times \DET_{W_b} \too \mathbb{C}, \hskip 5mm
  g_{\phi}([\a,\la],[\b,\mu]) = \ol{\la}\mu det_{F}(\a^{*}P_{W}\b),
\end{equation}
where $\a :W_b\to W$, (resp. $\b :W_b \to W^{'}$, is
antiholomorphic (resp. holomorphic) and antilinear (resp. linear)
in the first (resp. second) variable. We then have the Pl\"ucker
embedding map
\begin{equation}\label{e:pem}
\DET_{W_b}-\{0\} \too \Ff_{W_b}
\end{equation}
which maps $[\omega,\la]\mapsto \ol{\la}\phi(\omega, \ . \ )$, or,
using the Segal definition of the determinant line
\begin{equation}\label{e:pl}
 [\a,\la]\longmapsto \longmapsto \psi_{[\a,\la]}
 \end{equation}
defined for $[\a,\la] = [P_W \a,\la]\in \DET(W_b,W)$ and $\xi :
W_b\to W^{'}$ by
\begin{equation}\label{e:eval}
  \psi_{[\a,\la]}(\xi) = \ol{\la} det_{F}(\a^{*}\circ \xi) =
  \ol{\la} det_{F}(\a^{*}\circ P_{W}\circ \xi).
\end{equation}
 Notice that
\begin{equation}\label{e:detvac}
det(id_{W_b}) \longmapsto \nu_{W_b},
\end{equation}
where $id_{W_b} := P_{W_{b},W_{b}}$. The map \eqref{e:pem} thus
defines a projective embedding $Gr_b\to \Ff_W$

The map \eqref{e:phi2} restricted to a linear map $\DET_{W_b
}\times \DET_{W_b}\to \mathbb{C}$, defines a canonical metric on
$\DET_{W_b}$ by $\|[\a,\la]\|^{2} = |\la|^{2}det_{F}(\a^{*}\a)$,
and globally, via \eqref{e:isom3}, we get the {\em canonical
metric} of \cite{Sc98} on $\Det(\Df,\Pf)$
\begin{equation}\label{e:canmetric}
\|det \ D_{P_b}\|^{2} := g_{\phi}(S(P_b),S(P_b)) =
det_{F}(S(P_b)^{*}S(P_b)).
\end{equation}

On the other hand, we can use the map $\phi$ (or $g_{\phi}$) to
put a unitary structure on $\Ff_W$ with respect to which
\eqref{e:pem} is an isometry. To do that we use the fact that any
section in $\Ff_W$ can be written as a linear combination of the
$\psi_{[\a,\la]}$, and set
\begin{equation}\label{e:Fmetric}
  <\psi_{[\a,\la]},\psi_{[\b,\mu]}>_{W} =
  g_{\phi}([\a,\la],[\b,\mu]).
\end{equation}
In particular, the finite linear combinations of the sections
$\psi_{S},S\in \Ss$ are dense in $\Ff_W$ with respect to the
topology of uniform convergence on compact subsets, and one has
$$<\psi_{S},\psi_{S^{'}}> = \phi(\xi(S),\xi(S^{'})) =
\delta_{SS^{'}}.$$ Notice further the identities
\begin{equation}\label{e:iden}
<\nu_{W},\nu_{W}>_{W} =1 \hskip 5mm {\rm and} \hskip 5mm
<\psi_{[\a,\la]},\psi_{[\a,\la]}>_{W} = \|[\a,\la]\|^{2},
\end{equation}
the latter being the statement that \eqref{e:pem} is an isometry.
For further details see \cite{PrSe86} and \cite{Mi89}.

There is a different way of thinking about Fock spaces which is
perhaps more familiar to physicists, as an infinite-dimensional
exterior algebra (fermionic Fock space). Recall \cite{Mi89} that a
polarization $W$ of the Hilbert space $H_b$ fixes a representation
of the canonical anticommutation relations (CAR) in a Fock space
${\mathcal F}(H_b,W),$ whose only non-zero anticommutators are
\begin{equation}\label{e:CAR}
   a^*(v)a(u) + a(u)a^*(v) = <u,v>.
\end{equation}
The defining property of this irreducible representation is that
there is a vacuum vector $|W>$ with the property
\begin{equation}\label{e:CARvac}
  a(u)|W> = 0 = a^*(v)|W>  \text{ for all $u\in W, v\in
W^{\perp}$. }
\end{equation}
 One has
\begin{equation}\label{e:CARext}
  \Ff(H_b,W) = \wedge(W) \otimes \wedge((W^{\perp})^*) =
\sum_{d = q-p \in \mathbb{Z}}  \wedge^{p}(W) \otimes
\wedge^{q}((W^{\perp})^*).
\end{equation}
The vacuum $|W>$ is represented as the unit element in the
exterior algebra. For $u\in W,$ $a(u)$ corresponds to interior
multiplication by $u,$ the creation operator $a^*(u)$ is given by
exterior multiplication. For $u\in W^{\perp}$ the operator $a(u)$
(resp., $a^*(u)$) is given by exterior (resp., interior)
multiplication by $Ju;$ here $J: H\to H^*$ is the canonical
antilinear isomorphism from a complex Hilbert space to its dual.
The vacuum $|W>$ has then the characteristic property $a(u)|W> =
0, u\in W,$ and $ a^*(v)|W> = 0, v\in W^*.$.

If we choose a different $W'\in Gr_b$ then there is a complex
vacuum line in ${\mathcal F}(H_b,W)$ corresponding to the new
polarization $W'.$ The different vacuum lines parameterized by the
planes $W'$ form another realization for the determinant bundle
$\DET_W$ over $Gr_b$ as a subbundle of the trivial Fock bundle
with fibre $\Ff_W$. The Pl\"ucker embedding $\DET_W \to
\Ff(H_b,W)$ is defined by mapping $(\omega,\la) \in St_W\times
\mathbb{C}$ to $\la \sum_{S\in\Ss} det \omega_{S} \,\psi_S,$ where
$\omega_{S}$ is as before. A Hermitian metric on $\Ff(H_b,W)$ is
again defined by $<\psi_S, \psi_{S'}> = \delta_{SS^{'}}$. On the
other hand, the finite-dimensional matrix identity for $\a
:\mathbb{C}^m\to\mathbb{C}^n, \b :\mathbb{C}^n\to\mathbb{C}^m$
with $n\leq m$:
\begin{equation}\label{e:finiteab}
  det(\alpha\beta) = \sum_{(i)} det(\alpha(i)) det(\beta(i)),
\end{equation}
the sum being over all sequences $(i) = \{1\leq i_1<i_2 \dots i_n
\leq m\}$, with $\a(i)$ (resp. $\b(i)$) the matrix obtained from
$A$ (resp. $B$) by selecting the columns of $A$ (resp. rows of
$B$) labeled by $S$, implies the pairing of Fock space vectors
\begin{equation} \label{e:sumxi}
<\psi_{[\alpha,\lambda]}, \psi_{[\beta,\mu]}> = \sum_{S\in \Ss}
\psi_{\alpha, \lambda}(\xi(S))^* \psi_{[\beta,\mu]} (\xi(S)).
\end{equation}
The metrics so defined on the CAR construction $\Ff(H_b,W)$ and
the geometric construction $\Ff_{W}$ of the Fock space, then
correspond under the algebraic isomorphism defined by associating
to each section $\psi_{S}\in \Ff_{W}$ the vector $$a(e_{i_1})\dots
a(e_{i_p}) a^*(e_{j_1}) \dots a^*(e_{j_q})|W>\in \Ff(H_b,W),$$
where $i_1< i_2 <\dots i_p\leq 0$ is the set of negative indices
in the sequence $S$ and $0 < j_1 < j_2< \dots j_q$ is the set of
\it missing positive \rm indices, giving a dense inclusion $
\Ff(H_b,W) \too \Ff_{W}.$

\bigskip

Returning to the case of a family $\Df$ of Dirac-type operators
parameterized by a manifold $B$, if we are given a spectral
section $\Pf\in \Gg r(M/B)$ then we have the global version of the
above properties. Associated to $\Pf$ we have a Fock bundle
$\Ff_{\Ww} \to B$ and this is endowed with a unitary structure $<
\ , \ >_{\Pf}$, given on the fibre $\Ff_{W_b}$ by
\eqref{e:Fmetric}. The bundle $\Ff_{\Ww}$ has a distinguished
section, the {\em vacuum section} $\nu_\Pf = \nu_{\Ww}$, assigning
to $b\in B$ the vacuum vector $\nu_{W_b}$, with unit norm in the
fibres. Associated to the canonical Calderon section $P(\Df)$
defining the Fock bundle $\Ff_{\Kk}$ we then have a determinant
line bundle $\DET(\Df,\Pf)\cong\DET(\Sf(\Pf))$ and a `generalized
Pl\"ucker embedding'
\begin{equation}\label{e:genPlucker}
  \DET(\Df,\Pf)\cong\DET(\Sf(\Pf)) \too \Ff_{\Kk},
\end{equation}
corresponding to the view point on the parameter space $B$ as a
generalized Grassmannian. More generally, for any pair of spectral
sections $\Pf_1,\Pf_2$, there is a `generalized Pl\"ucker
embedding'
\begin{equation}\label{e:genPlucker2}
\DET(\Pf_1,\Pf_2) \too \Ff_{\Pf_1},
\end{equation}
 defined fibrewise by the embeddings
$\Det(W_{1,b},W_{2,b}) \hookrightarrow \DET_{W_{1,b}} \to
\Ff_{W_{1,b}}$, which  according to \eqref{e:pem} is the map
$[\a,\la]\mapsto \psi_{[\a,\la]}\in \Ff_{\Pf_1}$. So, a section of
the determinant bundle $\DET(\Pf_1,\Pf_2)$ defines a section of
the Fock bundle $\Ff_{\Pf_1}$. In particular, the vacuum section
is the image of the determinant section in the `trivial' case
\begin{equation}\label{e:genPluckervac}
\DET(\Pf_1,\Pf_1) \too \Ff_{\Pf_1},
\end{equation}
Associated to the family of Dirac operators $\Df$, we have a
canonical vacuum section $\nu_{\Kk}\in \Ff_{\Kk}$ with
$<\nu_{K_b},\nu_{K_b}>_{K_b} = 1$, and if we choose an external
spectral section $\Pf$, then via \eqref{e:genPlucker} we have a
canonical section $\psi_{\Kk,\Pf}$ of $\Ff_{\Kk}$ corresponding to
the determinant section $b\mapsto det(\Df_{P_b})\leftrightarrow
det(S(P_b))$ of  $\DET(\Df,\Pf)$, with
\begin{equation}\label{e:Fdetmetric}
<\psi_{\Kk,\Pf},\psi_{\Kk,\Pf}>_{K_b} =
\|det(S(P_{b}))\|^{2}_{\Cc}.
\end{equation}
That is, the generalized Pl\"ucker embedding \eqref{e:genPlucker}
is an isometry with respect to the canonical metric on
$\DET(\Df,\Pf)$. This follows by construction from \eqref{e:iden}.

As we already mentioned, as an abstract vector bundle the Fock
bundle is trivial. However, the non-triviality of the construction
lies in the (locally defined) physical vacuum subbundle defined by
the family of Hamiltonians. As an example,  assume that we have a
family of Dirac Hamiltonians parameterized by the set $\mathcal A$
of smooth vector potentials. Given a real number $\lambda$ we can
define $W_0(A)$ as the subspace of the boundary Hilbert space
corresponding to the spectral restriction $D_{Y,A} > \lambda$ for
the boundary Hamiltonian; $A\mapsto W_0(A)$ is a smooth Grassmann
section over the set $U_{\lambda} \subset \mathcal A$ of
Hamiltonians with $\lambda\notin Spec(D_{Y,A}).$ Let $A\mapsto
W_1(A)$ be a globally defined Grassmann section. For each $A\in
U_{\lambda}$ we have a well-defined vacuum line $|A>\in {\mathcal
F}_{W_1(A)}.$ This line is just the image of the determinant line
$\DET(W_1(A), W_0(A))$ with respect to the map \eqref{e:eval}. If
dim$\,Y=1$ the Grassmannian $Gr_A$ does not depend on the
parameter $A$ and we may take $W_1(A)$ as a constant section.
Anyway, the bundle of vacua over $U_{\lambda}$  can be identified
as the relative determinant bundle $\DET(W_1,W_0)$ and the
twisting of this bundle depends solely on the twisting of the
local section $A\mapsto W_0(A).$

\bigskip


\section{Construction of the FQFT}

In this section we utilize the facts presented in the previous
section to piece together a FQFT, generalized from the two
dimensional case proposed by Segal \cite{Seg90}. As in Section
1.1, the constructions are mathematical and do not refer to any
particular physical system. In the next section we explain how the
chiral anomaly and commutator anomaly arise in this context.

\subsection{Strategy}

We define a projective functor from a subcategory $\Cc_{d}$ of the
category of spin manifolds to the category $\Cc_{vect}$ of
$\mathbb{Z}$-graded vector spaces and linear maps, which factors
through the category $\Cc_{Gr}$ of {\em linear relations}:
\begin{eqnarray*}
\Cc_{d} & \too & \Cc_{Gr} \\
 &  \searrow & \downarrow \\
& & \Cc_{vect}
\end{eqnarray*}
The combination of these functors is the Fock functor defining the
FQFT.

\subsection{Projective representations of categories} By a
category $\Cc$ we mean a set $\Ob(\Cc)$ of elements called the
objects of $\Cc$, and for any two elements $a,b\in \Ob(\Cc)$ a set
$\Mor_{\Cc}(a,b)$ of morphisms $a\mto b$, such that for $a,b,c\in
\Ob(\Cc)$ there is a multiplication defined $$
\Mor_{\Cc}(a,b)\times\Mor_{\Cc}(b,c) \too \Mor_{\Cc}(a,c),\hskip
5mm (f_{a,b},f_{b,c})\mtoo f_{b,c}\circ f_{a,b}.$$

\noi The product is required to be associative, so that if
$f_{c,d}\in \Mor_{\Cc}(c,d)$, then $f_{c,d}(f_{b,c}f_{a,b}) =
(f_{c,d}f_{b,c})f_{a,b}$. One usually also asserts the existence
of an identity morphism $id_{b}\in \Mor_{\Cc}(b,b)$ which
satisfies $id_{b}\circ f_{a,b} = f_{a,b}$ and $f_{b,c}\circ id_{b}
= f_{b,c}$.

A (covariant) {\em functor} $\Psi$ from a category $\Cc$ to a
category $\Cc^{'}$ means a map $\Psi : \Ob(\Cc) \to \Ob(\Cc^{'})$
and for each pair $a,b\in \Ob(\Cc)$ a map
$\Psi_{a,b}:\Mor_{\Cc}(a,b)\to\Mor_{\Cc^{'}}(\Psi(a),\Psi(b))$
such that
\begin{equation}\label{e:functor1}
\Psi_{a,c}(f_{b,c}f_{a,b}) =
\Psi_{b,c}(f_{b,c})\Psi_{a,b}(f_{a,b}).
\end{equation}
If $\Cc^{'}$ is the category of vector spaces and linear maps,
then $\Psi$ is a {\em representation} of the category $\Cc$.

A classical result of Wigner tells us that in quantum systems we
must content ourselves with projective representations of symmetry
groups. Similarly, with the Fock functor we have to consider
projective category representations. This means that there is
essentially a scalar ambiguity in the map $\Psi_{a,b}$, so that
\eqref{e:functor1} is replaced by
\begin{equation}\label{e:functor2}
\Psi_{a,c}(f_{b,c}f_{a,b}) =
c(f_{b,c},f_{a,b})\Psi_{b,c}(f_{b,c})\Psi_{a,b}(f_{a,b}),
\end{equation}
where the `cocycle' $c(f_{b,c},f_{a,b})$ takes values in
$\mathbb{C}-\{ 0\}$. To explain the meaning here of `essentially',
recall that a projective representation of a group $G$ is a true
representation of a extension group $\hat{G}$ of $G$ by
$\mathbb{C}^{\times}$. The group $\hat{G}$ forms a
$\mathbb{C}^{\times}$ bundle over $G$ whose Lie algebra cocycle is
the first Chern class of the associated line bundle. Equivalently,
$\hat{G}$ is defined by assigning to each $g\in G$ a complex line
$L_{g}$ such that $L_{g_1 g_2} = L_{g_1}\otimes L_{g_2}$. (A
well-known instance of this occurs for loop groups, see Example
4.1 below, and more generally we shall in Section 4 give the gauge
group representations for a Yang-Mills action functional a similar
description.) Likewise, a projective representation of a category
is a true representation of an extension category $\hat{\Cc}$
constructed by assigning to each $f_{a,b}\in \Mor_{\Cc}(a,b)$ a
complex line $L_{f_{a,b}}$, and given $f_{b,c}\in
\Mor_{\Cc}(b,c)$, an identification
\begin{equation}\label{e:projL}
L_{f_{a,b}} \otimes L_{f_{b,c}} \too  L_{f_{a,b}f_{b,c}} ,
\end{equation}
which is associative in the natural sense. One then has
\begin{equation}\label{e:projC}
\Ob(\hat{\Cc}) = \Ob(\Cc),\hskip 5mm \Mor_{\hat{\Cc}}(a,b) =
\{(f,\la) \ | \ f \in \Mor_{\Cc}(a,b), \  \la \in L_f\}.
\end{equation}

\subsection{The category $\Cc_{d}$}

An element of $\Ob(\Cc_d)$  is a pair $(\Yy,W)$, where $\Yy =
(Y,g_Y, S_Y \otimes \xi_Y)$, with $Y$ is a closed, smooth and
oriented d-dimensional spin manifold, $g_Y$ a Riemannian metric on
$Y$, $S_Y$ a spinor bundle over $Y$, $\xi_Y$ a Hermitian bundle
over $Y$ with compatible gauge connection, and $W$ is an
admissible polarization of the `one-particle' Hilbert space $H_Y =
W \oplus W^{\perp}$ to a pair of closed infinite-dimensional
subspaces. Here $H_Y = L^2(Y, S_Y\otimes \xi_Y)$ and admissible
means that $P_W \in Gr_Y$, where $P_W$ is the orthogonal
projection onto $W$ and $Gr_Y$ is the Hilbert-Schmidt Grassmannian
defined with respect to the energy polarization $H_Y =H^+\oplus
H^-$ into positive, resp. negative, energies of the Dirac operator
$D_Y.$

Let $(\Yy_i,W_i) \in \Ob(\Cc_d), \ i=1,2,$ where $\Yy_i =
(Y_i,g_{Y_i}, S_{Y_i} \otimes \xi_{Y_i})$. An element of
$\Mor_{\Cc_d}((\Yy_1,W_1),(\Yy_2,W_2))$ is a triple $\Xx = (X,g_X,
S_X \otimes \xi_X)$, where $X$ smooth and oriented
(d+1)-dimensional spin manifold with boundary $\dd X = Y_1 \sqcup
Y_2$, $g_X$ a Riemannian metric on $X$ with $(g_X)_{|Y_i} =
g_{Y_i}$, $S_X$ a spinor bundle and $\xi_X$ a Hermitian bundle
over $X$ with compatible gauge connection, such that $(S_X\otimes
\xi_X)_{|Y_i} \cong S_Y \otimes \xi_{Y_i}$ and the connections
metrics correspond under the isomorphism. We refer to $\Xx$ as a
{\em geometric bordism} from $\Yy_1$ to $\Yy_2$. We assume that:
\begin{itemize}
  \item In a collar neighbourhood of the boundary $U = U_1 \sqcup U_2$, where
   $U_i = ([0,1]\times Y_i)$ the geometry of all metrics, connections is
 a product. Recall this means that near the boundary the metric
becomes the product of the standard metric on the real axis and
the boundary metric. Similarly, the gauge connection approaches
smoothly the connection on the boundary such that at the boundary
all the normal derivatives vanish. Thus $\xi_{X|U_i}$ is a
pull-back of the boundary bundle $(\xi_{Y_i})$, and similarly all
metrics, connections, etc are pull-backs of their boundary
counterparts, so $g_{X|U_i} = du^2 + g_{Y_i}$ etc.
  \item The orientation on the 'ingoing' boundary $Y_1$ is assumed to be
induced by the orientation of $X$ and the inward directed normal
vector field on the boundary, whereas for the 'outgoing' boundary
$Y_2$ the orientation is fixed by the outward directed normal
vector field.
\end{itemize}

 For notational brevity we may write $ S_{i} := S_{Y_i} \otimes
 \xi_{Y_i}, \ g_i := g_{Y_i}$ etc, and $S_{1,2} := S_X \otimes
 \xi_X, \ g_{1,2} := g_{X}$ etc, in the following.

We augment $\Cc_d$ by including the empty set
$\emptyset\in\Ob(\Cc_d)$, and for each $(\Yy,W)\in \\Ob(\Cc_d)$ we
also allow $\emptyset := id_b$ as an element of
$\Mor_{\Cc_d}((\Yy,W),(\Yy,W))$. In particular, a geometric
bordism $\Xx\in \Mor_{\Cc_d}(\emptyset,(\Yy,W))$ means
$d+1$-dimensional manifold $X$ with boundary $Y$ (plus bundles,
connections etc). Thus a morphism in $\Cc_d$ may have
disconnected, connected, or empty (i.e. $X$ is closed) boundary,
according as $Y$ is disconnected, connected or empty.

For $(\Yy_i,W_i)\in\Ob(\Cc_d), \ i=1,2,3$, there is an associative
product map
\begin{equation}\label{e:Cdmult1}
\Mor_{\Cc}((\Yy_1,W_1),(\Yy_2,W_2))\times\Mor_{\Cc}((\Yy_2,W_2),(\Yy_3,W_3))
\too \Mor_{\Cc}((\Yy_1,W_1),(\Yy_3,W_3))
\end{equation}
taking a pair $(\Xx_{1,2},\Xx_{2,3})$ to the geometric bordism
\begin{equation}\label{e:Cdmult2}
\Xx_{1,2}\cup_{Y_2}\Xx_{2,3} = (X_{1,2}\cup_{Y_2}X_{2,3}
,g_{1,2}\cup g_{2,3} , S_{1,2}\cup_{\sigma}S_{2,3}).
\end{equation}
This `sewing together' of bundles is defined in the usual way.
Briefly, the collar neighbourhood $U_{r} = [0,1)\times Y_2$ of the
boundary of $Y_2$ in $X_{2,3}$ is a copy of the collar
neighbourhood $U_{l} = (-1,0] \times Y_2$ of the boundary of $Y_2$
in $X_{1,2}$ but with orientation reversed. Hence we may glue
together the manifolds $X_{1,2}$ and $X_{2,3}$ along $Y_2$ to get
the `doubled' manifold $X_{1,2}\cup_{Y_2}X_{2,3}$ with a tubular
neighbourhood of the partition $Y_2$ which we may parameterize as
$U = (-1,1)\times Y_2$. Associated to the geometric data we have
Dirac operators $D_{1,2}$ and $D_{2,3}$ acting respectively on
sections of the Clifford bundles  $S_{1,2}$ and $S_{2,3}$. Over
$U_l$ the operator $D_{1,2}$ takes the product form $\sigma
(\dd/\dd u + D_{Y_2})$, because of the change of orientation
$(D_{2,3})_{|U_r} = (\dd/\dd v + D_{Y_2})\sigma\ii$. Over $Y_2$ we
construct $S_{1,2}\cup_{\sigma} S_{2,3}$ by gluing $S_{1,2}$ to
$S_{2,3}$ via the unitary isomorphism $\sigma$, identifying $s\in
(S_{1,2})_{|Y}$ with $\sigma s\in (S_{2,3})_{|Y}$. (Thus for the
case of chiral spinors the isomorphism $\sigma$ takes positive to
negative spinors.) A section of $S_{1,2}\cup_{\sigma}S_{2,3}$ is a
pair $(\psi,\phi)$ with $\psi$ (resp. $\phi$) is a smooth section
of $S_{1,2}$ (resp. of $S_{2,3}$) such that the normal derivatives
of all orders match-up: $\frac{\dd^{k}}{\dd u^{k}}\psi(0,y) =
(-1)^{k}\sigma(y) \frac{\dd^{k}}{\dd u^{k}}\phi(0,y)$. We then
have the `doubled' Dirac-type operator $(D_{1,2}\cup
D_{2,3})(\psi,\phi) = (D_{1,2}\psi, D_{2,3}\phi)$ acting on
$\Ci(X_{1,2}\cup_{Y_2}X_{2,3},S_{1,2}\cup_{\sigma}S_{2,3})$, which
is well-defined since from the product form \eqref{e:Du} it can be
easily checked that $D_{1,2}\psi$ and $D_{2,3}\phi$ match up at
the boundary.

\bigskip
\subsection{The category $\Cc_{Gr}$}

An element of $\Ob(\Cc_{Gr})$ is a pair $(H,W)$ with $H$ a Hilbert
space, and $W$ a polarization of $H$ into a pair of closed
orthogonal infinite-dimensional subspaces $H = W \oplus
W^{\perp}$. A morphism $(E,\e)\in
\Mor_{\Cc_{Gr}}((H_1,W_{1}\pp),(H_2,W_2))$ is a closed subspace
$E\subset H_1 \oplus H_2$ such that $P_E - P_{W_{1}^{\perp}\oplus
W_{2}}$ is a Hilbert-Schmidt operator, where $P_E,
P_{W_{1}\pp\oplus W_{2}}$ are the orthogonal projections with
range $E,W_{1}^{\perp}\oplus W_{2} $ respectively, along with an
element $\e$ of the relative determinant line $\Det(W_{1}\pp\oplus
W_{2},E)$. (It is convenient here to use the `reverse'
polarization $W_{1}\pp$ of $H_1$ in order to account for boundary
orientations later on, see below). Thus there is an identification
\begin{equation}\label{e:Grmor}
\Mor_{\Cc_{Gr}}((H_1,W_{1}\pp),(H_2,W_2)) = \DET_{W_{1}\pp\oplus
W_{2}},
\end{equation}
where the right-side is the determinant line  {\em bundle} based
at $ W_{1}\pp\oplus W_{2}$ over the trace-class Grassmannian
$Gr(\ol{H}_1 \oplus H_2)$, where $\ol{H}_1$ serves to remind us
that we are considering the reverse polarization $W_{1}\pp$; we
may write $Gr (\ol{H}_1 \oplus H_2) = Gr(\ol{H}_1 \oplus H_2,
W_{1}\pp\oplus W_2)$ and $\DET_{W_{1}\pp\oplus W_{2}} =
\DET_{W_{1}\pp\oplus W_{2}}(\ol{H}_1 \oplus H_2)$ if we wish to
emphasize the polarization. We also allow $\O\in\Ob(\Cc_{Gr})$ as
an object, and define
\begin{equation}\label{e:Grmor2}
\Mor_{\Cc_{Gr}}(\O,(H,W)) = \DET_{W}(H),\hskip 5mm
\Mor_{\Cc_{Gr}}((\ol{H},W\pp),\O) = \DET_{W\pp}(\ol{H})
\end{equation}
$$\Mor_{\Cc_{Gr}}(\O,\O) = \mathbb{C}.$$

To define the product of morphisms in $\Cc_{Gr}$, first recall
when $H_0 \neq\O \neq H_2$, from the `category of linear
relations', the `join' product rule
\begin{equation}\label{e:Grmult}
Gr(\ol{H}_0 \oplus H_1,W_{0}\pp\oplus W_1)  \times Gr(\ol{H}_1
\oplus H_2, \wt{W}_{1}\pp\oplus W_2) \too Gr (\ol{H}_0 \oplus
H_2,W_{0}\pp\oplus W_2),
\end{equation}
$$(E_{01},E_{12}) \mtoo E_{01}*E_{12},$$ where $\wt{W}_1\in
Gr(H_1,W_1)$, defined by $$E_{01}* E_{12} = \{(u,v)\in H_{0}
\oplus H_{2}\ | \ \exists \ w\in H_1 \ \text{such \ that} \
(u,w)\in E_{12}, (w,v)\in E_{23} \}.$$ The join is a generalized
composition law of graphs of linear operators, but here the
morphisms $E$ are not in general everywhere defined, but $dom(E) =
range(P_{H_1}P_E :E\to H_1)$, and may also be `multi-valued'. The
composition may therefore be discontinous. From \cite{Seg90} we
recall that for continuity one requires that: (i) the map
$E_{01}\oplus E_{12} \to H_1, \ ((u,w),(w^{'},v)\mto w - w^{'}$ is
surjective, and (ii) $E_{01}\oplus E_{12} \to H_0\oplus H_1 \oplus
H_2, \ ((u,w),(w^{'},v)\mto (u, w - w^{'},v)$ is injective. The
crucial fact is the following:
\begin{prop}\label{p:detpairing}
With the above notation, when $(\ol{H_0},W_{0}\pp) = \O =
(H_2,W_{2})$ there is a canonical pairing, linear and holomorphic
on the fibres in the first and second variables,
\begin{equation}\label{e:detpairing2}
\k : \DET_{W_1}\times \DET_{\wt{W}_{1}\pp}\too \Det(W_1,\wt{W}_1).
\end{equation}
If $W_1 = \wt{W}_{1}$, then
\begin{equation}\label{e:detpairing2a}
\k : \DET_{W_1}\times \DET_{\wt{W}_{1}\pp}\too \mathbb{C}.
\end{equation}
  More generally, if (i) and (ii) hold,
then one has such a pairing
\begin{equation}\label{e:detpairing1}
\k : \DET_{W_{0}\pp\oplus W_{1}}(\ol{H_0} \oplus H_1)\times
\DET_{\wt{W}_{1}\pp\oplus W_{2}}(\ol{H}_1 \oplus H_2) \too
\end{equation}
$$\DET_{W_{0}\pp\oplus W_{2}}(\ol{H}_0 \oplus H_2)\otimes
\Det(W_1,\wt{W}_1),$$ which respects the join multiplication: on
each fibre
 \begin{equation}\label{e:detpairing3}
\k : \Det(W_{0}\pp\oplus W_{1},E_{01})\times
\DET(\wt{W}_{1}\pp\oplus W_{2},E_{12}) \too
\end{equation}
$$\Det(W_{0}\pp\oplus W_{2},E_{01}*E_{12})\otimes
\Det(W_1,\wt{W}_1).$$ (Here the second factor on the right-side of
\eqref{e:detpairing1} denotes the trivial bundle with fibre
$\Det(W_1,\wt{W}_1)$.) If $W_1 = \wt{W}_{1}$, then
\begin{equation}\label{e:detpairing3a}
\k : \DET_{W_{0}\pp\oplus W_{1}}(\ol{H_0} \oplus H_1)\times
\DET_{W_{1}\pp\oplus W_{2}}(\ol{H}_1 \oplus H_2) \too
\DET_{W_{0}\pp\oplus W_{2}}(\ol{H}_0 \oplus H_2).
\end{equation}
 \end{prop}
\begin{proof}
As before, we denote by $P_{W,W'}$ the orthogonal projection onto
$W$ restricted to the
 subspace $W'.$
 Given $E\in Gr(H_1,W_1),E^{'}\in Gr(\ol{H}_1,\wt{W}_{1}\pp) =
Gr(\ol{H}_1,W_{1}\pp)$, we can represent elements $\e \in
\Det(W_1,E)$ and $\d \in \Det(\wt{W}_{1}\pp,E^{'})$ as the
determinant elements of linear operators $a_{\e}:W_1\to E$ and
$b_{\d}:\wt{W}_{1}\pp\to E^{'}$ with $a_{\e}-P_{E,W_1}$ and
$b_{\d}-P_{E^{'},\wt{W}_{1}\pp}$ of trace-class; consequently, the
operators $P_{W_1} a_{\e} - id_{W_1}$ and $P_{\wt{W}_{1}\pp}
b_{\d}- id_{\wt{W}_{1}\pp}$ are trace-class, too. We define
\begin{equation}\label{e:detpairing4}
  \k : \Det(W_1,E)\times \Det(\wt{W}_{1}\pp,E^{'})\too
  \Det(W_1,\wt{W}_1),
\end{equation}
by
\begin{equation}\label{e:detpairing5}
\k(\e,\d) = det(P_1 a_{\e} +
\wt{P}_{1}^{\perp}b_{\d})\in\Det(W_1\oplus \wt{W}_{1}\pp,
H_1)\cong \Det(W_1,\wt{W}_1),
\end{equation}
where $P_1,\wt{P}_{1}$ are the projections on $W_1,\wt{W}_1$, and
$\Det(W_1\oplus \wt{W}_{1}\pp, H_1)$ is the determinant line of $
P_1  + \wt{P}_{1}^{\perp}:W_1\oplus \wt{W}_{1}\pp\to H_1$. That
this operator differs from  $P_1 a_{\e} +
\wt{P}_{1}^{\perp}b_{\d}$ by an operator of trace-class (in order
that \eqref{e:detpairing4} be well-defined) follows immediately
from that fact that the operators $P_{W_1} a_{\e} - id_{W_1}$ and
$P_{\wt{W}_{1}\pp} b_{\d}- id_{\wt{W}_{1}\pp}$ are trace-class.

The canonical isomorphism on the right-side of
\eqref{e:detpairing5} is expressed via the diagram of commutative
maps with exact rows and Fredholm columns $$\begin{CD} 0 @>>>
W_{1} @>>> W_{1}\oplus \wt{W}_{1}\pp @>>> \wt{W}_{1}\pp @>>> 0\\
  @.  @VV{\wt{P}_1 P_1 }V @VV{\wt{P}_1 P_1  + \wt{P}_{1}^{\perp}}V  @VV{id}V \\
0   @>>> \wt{W}_{1}@>>>  H_1  @>>> \wt{W}_{1}\pp @>>> 0
\end{CD}
$$ where the horizontal maps  are the obvious ones. We know from
\cite{Seg90,Sc98} that such a diagram defines an isomorphism
between the determinant line of the centre map with the tensor
product of the lines defined by the outer columns, mapping the
determinant elements to each other. Hence since $\Det(id) =
\mathbb{C}$ canonically, the isomorphism follows, and in
particular with $E = W_1$ and $E^{'} = \wt{W}_1\pp$ we have
\begin{equation}\label{e:detpairing6}
\k(det(id_{W_1}),det( id_{\wt{W}_1\pp})) = det(P_{\wt{W}_1, W_1}),
\end{equation}
where $id_{W}=P_{W,W}$, which will be a relevant fact later in
this Section.

For the general case \eqref{e:detpairing1}, suppose initially that
$W_1 = \wt{W}_1$ and choose $\e \in \Det(W_{0}\pp\oplus
W_{1},E_{01})$ and $\d \in \DET(W_{1}\pp\oplus W_{2},E_{12})$
identified with the determinant elements of linear operators
$a_{\e}:W_{0}\pp\oplus W_{1}\to E_{01}$ and $b_{\d}:W_{1}\pp\oplus
W_{2}\to E_{12}$. Define
\begin{equation}\label{e:detpairing7}
\k_1 : \Det(W_{0}\pp\oplus W_{1},E_{01})\times \DET(W_{1}\pp\oplus
W_{2},E_{12}) \too
\end{equation}
$$ \DET(W_{0}\pp\oplus H_1\oplus W_{2},E_{01}\oplus E_{12}),$$
$$\k_1 (\e,\d) = det(a_{\e}\oplus b_{\d}).$$ On the other hand,
from \cite{Seg90}, conditions (i) and (ii) mean that there is an
exact sequence
\begin{equation}\label{e:exact}
0 \too E_{01}* E_{12} \too E_{01}\oplus  E_{12} \too H_1 \too 0,
\end{equation}
and this fits into the commutative diagram with Fredholm columns
$$\begin{CD} 0 @>>> E_{01}* E_{12} @>>> E_{01}\oplus E_{12} @>>>
H_{1} @>>> 0\\
  @.  @VV{P_{W_{0}\pp\oplus W_{2}}P_{E_{01}* E_{12}}}V @VV{G}V
  @VV{id}V \\
0   @>>>  W_{0}\pp\oplus W_2 @>>>  W_{0}\pp\oplus H_1 \oplus W_2
@>>>  H_1 @>>> 0
\end{CD}
$$ where we modify \eqref{e:exact} by composing the injection
$E_{01}* E_{12} \too E_{01}\oplus  E_{12}$ with the involution
$((u,w),(w^{'},v)\to ((u,w),(-w^{'},v)$, and the following
surjection to $((u,w),(w^{'},v)\mto (u, w + w^{'},v)$, while the
lower maps are again the obvious ones. The central column is
$$G(\xi,\eta) = (P_{W_{0}\pp} P_{H_{0}}\xi, P_{H_1}\xi +
P_{H_1}\eta, P_{W_{2}}P_{H_1}\eta).$$

Because $P_{H_{1}}P_{E_{01}} - P_{W_{1}}P_{H_{1}}P_{E_{01}} =
P_{H_{1}}(P_{E_{01}} -P_{W_{0}\pp\oplus W_{1}})P_{E_{01}} $ and
$P_{H_{1}}P_{E_{12}} -
P_{W_{1}\pp}P_{H_{1}}P_{E_{12}}=P_{H_1}(P_{E_{12}}-P_{W_{1}\pp\oplus
W_2}) P_{E_{12}}$ are by trace-class, the operators $G$ and
$G_{W_1}$, where

$$G_{W_1}(\xi,\eta) = (P_{W_{0}}\pp P_{H_{0}}\xi,
P_{W_1}P_{H_1}\xi + P_{W_1}\pp P_{H_1}\eta,
P_{W_{2}}P_{H_1}\eta),$$ differ by only trace-class operators and
so $\Det(G) = \Det(G_W) =  \DET(E_{01}\oplus E_{12},W_{0}\pp\oplus
H_1\oplus W_{2})$, while from the diagram we have $ \Det(G ) \cong
\Det(E_{01}* E_{12},W_{0}\pp\oplus W_{2})$. Thus by duality (i.e.
take adjoints in the above diagrams, reversing the order of the
columns and rows and the direction of the arrows) we have a
canonical isomorphism $\Det(W_{0}\pp\oplus W_{2},E_{01}* E_{12})
\cong  \DET(W_{0}\pp\oplus H_1\oplus W_{2}, E_{01}\oplus E_{12})$,
and so composition with $\k_1$ completes the proof of
\eqref{e:detpairing1} in the case $W_1 = \wt{W}_1$. In the general
case, replace $H_1$ in \eqref{e:detpairing7} and the lower row of
the commutative diagram by $W_1\oplus\wt{W}_{1}\pp$ and repeat the
argument used in the proof of \eqref{e:detpairing2}. Finally, we
note for later reference that in the `vacuum case' $E_{01} =
W_{0}\pp\oplus W_1 $ and $E_{12} = W_{1}\pp\oplus W_2$ one has
$E_{01}*E_{12} = W_{0}\pp\oplus W_2 $ and
\begin{equation}\label{e:detpairing8}
\k(det(id_{E_{01}}),det(id_{E_{12}} ) = det(id_{E_{01}*E_{12}} ).
\end{equation}
\end{proof}

>From \eqref{e:detpairing3a} and the identification \eqref{e:Grmor}
we now have a canonical multiplication
\begin{equation}\label{e:Grmult1}
\Mor_{\Cc_{Gr}}((H_0,W_{0}\pp),(H_1,W_1)) \times
\Mor_{\Cc_{Gr}}((H_1,W_{1}\pp),(H_2,W_2))
\end{equation}
$$ \too \Mor_{\Cc_{Gr}}((H_0,W_{0}\pp),(H_2,W_2)),$$
$$(E_{0,1},\e), (E_{1,2},\d)) \mtoo (E_{0,1} * E_{1,2},\e *\d),$$
where $\e *\d := \k(\e,\d)$ if (i) and (ii) hold, and  $\e *\d :=
0 $ otherwise. In particular,
\begin{equation}\label{e:GrmultO}
\Mor_{\Cc_{Gr}}(\O,(H_1,W_1)) \times
\Mor_{\Cc_{Gr}}((H_1,W_{1}\pp),\O) \too \Mor_{\Cc_{Gr}}(\O,\O),
\end{equation}
is precisely equation \eqref{e:detpairing2}.
\bigskip

\subsection{The projective functor $\Cc_d\to\Cc_{Gr}$}

We define a projective functor $\Psi : \Cc_d\to\Cc_{Gr}$ as
follows. We have
\begin{equation}\label{e:func1}
  \Psi : \Ob(\Cc_d) \too \Ob(\Cc_{Gr}), \hskip 5mm (\Yy,W) \mtoo
  (H_Y,W),
\end{equation}
where as before $H_Y = L^2(Y, S_Y\otimes \xi_Y)$ and $W$ is an
admissible polarization. While for $(\Yy_1,W_1),(\Yy_2,W_2)\in
\Ob(\Cc_d) $
\begin{equation}\label{e:func2}
  \Psi : \Mor_{\hat{\Cc_d}}((\Yy_1,W_1),(\Yy_2,W_2))
  \too \Mor_{\hat{\Cc_{Gr}}}((\ol{H}_{Y_1},W_1\pp),(H_{Y_2},W_2))
\end{equation}
$$\Xx \mtoo (K_{12}, \e),$$ where $K_{12}\subset H_{Y_1}\oplus
H_{Y_2}$ is the Calderon subspace of boundary `traces' of
solutions to the Dirac operator $D^{1,2}$ over $X$ defined by the
geometric data in $\Xx$, and $\e\in \Det(W_{1}\pp\oplus
W_2,K_{12}) \cong \Det(D^{1,2}_{P_{W_{1}\pp,W_2}})^{*}$. Taking
into account that $Y_1$ is an incoming boundary, we have
$K_{12}\in Gr(\ol{H}_{Y_1}\oplus H_{Y_2}; W_{1}\pp\oplus W_2)$ (in
fact, an element of the `smooth Grassmannian'). The choice needed
of the element $\e$ means that $\Psi$ is a true functor
$\widehat{\Cc_d}\to\Cc_{Gr}$, where $\widehat{\Cc_d}$ is the
extension category of $\Cc_d$ whose objects are the same as
$\Cc_d$, and
\begin{equation}\label{e:projCd}
\Mor_{\widehat{\Cc_d}}(\Yy_1,\Yy_2) = \{(\Xx,z) \ | \ \Xx \in
\Mor_{\Cc_d}(\Yy_1\Yy_2), \ \e\in \Det(W_{1}\pp\oplus W_2,K_{12})
\}.
\end{equation}
For a closed geometric bordism $\Xx\in\Mor_{\Cc_d}(\O,\O)$ we set
\begin{equation}\label{e:projCd2}
\Mor_{\widehat{\Cc_d}}(\O,\O) = \Det(D_{X}),
\end{equation}
the projectivity of the functor in this case corresponds to a
choice of generator $\Det(D_{X})\cong \mathbb{C}=
\Mor_{\Cc_{Gr}}(\O,\O)$.

To see the functor respects the product rules in each category, it
is enough to show that $K_{01}*K_{12}$ is the Calderon subspace of
the operator $D^{0,1}\cup D^{1,2}$, i.e. $K(D^{0,1}\cup D^{1,2}) =
K(D^{0,1})*K(D^{1,2})$ defined by morphisms $\Xx_{0,1},
\Xx_{1,2}$. This, however, is immediate from the definition of
$D^{0,1}\cup D^{1,2}$, and the fact that given $\psi\in \Ker \
D^{0,1}, \phi\in \Ker \ D^{1,2}$ it is enough for their boundary
values to match up in order to get an element of $\Ker \
(D^{0,1}\cup D^{1,2})$. That in turn follows because the product
geometry in the collar neighbourhood $U$ of the outgoing boundary
$Y_1$ of $X_{0,1}$ implies that $\psi$ has the form $\psi(u,y) =
\sum_{k}e^{-\la_{k}u}\psi_{k}(0)e_{k}(y)$, where $\{\la_k,e_k\}$
is a spectral resolution of $H_{Y_1}$ defined by the boundary
Dirac operator. (To be quite correct, we should also include the
identification by the boundary isomorphism $\sigma (y)$ in the
definition of the join $K_{01}*K_{12}$, but this introduces no new
phenomena.) Thus the requirement \eqref{e:projL} for the rule
$\Xx\mto  \Det(W_{1}\pp\oplus W_2,K_{12})$ to define a projective
extension of $\Cc_{Gr}$, is precisely \eqref{e:detpairing3a} of
\propref{p:detpairing}.

\subsection{The functor $\Cc_{Gr}\to \Cc_{vect}$}

The functor $\Phi$ from the category $\Cc_{Gr}$ to the category
$\Cc_{vect}$ of ($\mathbb{Z}$-graded) vector spaces and linear
maps, is defined on objects of $\Cc_{Gr}$ by
\begin{equation}\label{e:F1}
 (H,W) \too \Ff_W = \Ff_{W}(H) \hskip 5mm
(\ol{H},W\pp) \too \Ff_{W\pp} = \Ff_{W\pp}(\ol{H})
\end{equation}
$$\O \too \mathbb{C},$$ Thus $\Phi$ takes a polarized vector space
to the Fock space defined by the polarization, and $\Ff_{\O} =
\mathbb{C}$ is by {\it fiat}. Here $\Ff_{W\pp} =
\Ff_{W\pp}(\ol{H}) := \Gamma_{hol}(Gr(\ol{H}),\DET_{W\pp})$ is the
Fock space associated with the reverse polarization.

$\Phi$ is defined on morphisms as follows. From \eqref{e:Grmor},
an element of $\Mor_{\Cc_{Gr}}((H_1,W_{1}\pp),(H_2,W_2))$  is the
same thing as an element $\e \in \DET_{W_{1}\pp\oplus W_{2}}$,
which we may think of as the pair $(E,\e)$ where
$\e\in\Det(W_{1}\pp\oplus W_{2},E)$. By the Pl\"ucker embedding
\eqref{e:pem} this gives us a canonical vector
\begin{equation}\label{e:Pe}
  \phi_{\e}\in \Ff_{W_{1}\pp\oplus
W_{2}}(\ol{H}_1\oplus H_2) \cong \Ff_{W_{1}\pp}(\ol{H}_1)\otimes
\Ff_{W_{2}}(H_2).
\end{equation}
The isomorphism is immediate from \eqref{e:CARext} and
$\Ff_{\Ww}(H) = \ol{\Ff}(H,W)$, the completion of $\Ff(H,W)$.

To proceed we need the following facts, generalizing eqn.(8.10) of
\cite{Seg90}:

\begin{prop}\label{p:Fockpairing}
The determinant bundle pairing $\k$ of \propref{p:detpairing}
defines a canonical Fock space pairing
\begin{equation}\label{e:Fockpairing1}
( \, \ ) : \Ff_{W_1}\times \Ff_{\wt{W}_{1}\pp}\too
 \Det(W_1,\wt{W}_1),
\end{equation}
with
\begin{equation}\label{e:Fockpairing2}
( \nu_{W_1}, \nu_{\wt{W}_{1}\pp}) = det(P_{\wt{W}_{1},W_1}).
\end{equation}
 If $W_1 = \wt{W}_{1}$, this becomes
\begin{equation}\label{e:Fockpairing3}
( \, \ ) : \Ff_{W_1}\times \Ff_{\wt{W}_{1}\pp}\too \mathbb{C},
\hskip 5mm ( \nu_{W_1}, \nu_{W_{1}\pp}) = 1.
\end{equation}
More generally,  $\k$ defines a pairing
\begin{equation}\label{e:Fockpairing4}
\Ff_{W_{0}\pp\oplus W_{1}}(\ol{H}_0\oplus H_1)\times
\Ff_{\wt{W}_{1}\pp\oplus W_{2}}(\ol{H}_1 \oplus H_2) \too
\Ff_{W_{0}\pp\oplus W_{2}}(\ol{H}_0 \oplus H_2)\otimes
\Det(W_1,\wt{W}_1),
\end{equation}
with
 \begin{equation}\label{e:Fockpairing5a}
( \phi_{\e}, \phi_{\d})  = \phi_{\k(\e,\d)}.
\end{equation}
In particular,
\begin{equation}\label{e:Fockpairing5}
(\nu_{W_{0}\pp\oplus W_{1}}, \nu_{\wt{W}_{1}\pp\oplus W_{2}})  =
\nu_{W_{0}\pp\oplus W_{2}}\otimes det(P_{W_2, W_0}).
\end{equation} \end{prop}
\begin{proof}
First notice that in the finite-dimensional case there is a
natural isomorphism between the Fock space (the exterior algebra)
and its dual defined by the pairing
$\wedge^{k}H\times\wedge^{n-k}H\to \Det(H), \
(\lambda_1,\lambda_2)\mapsto \lambda_1 \wedge \lambda_2,$ while in
the infinite-dimensional case the pairing using the CAR
construction follows directly from the definition ${\mathcal
F}(H,W)= \wedge(W) \otimes \wedge((W^{\perp})^*).$ For the
geometric Fock space $\Ff_{W}$, the construction of the pairing
from the determinant bundle pairing  $\k$ on $\DET_W \times
\DET_{W\pp}$ is entirely analogous to the construction of the
inner-product $< \ , \ >_{W}$ on $\Ff_W$
 from the determinant bundle pairing
 $g_{\phi}$ on $\DET_W \times \DET_{W}$ in equation \eqref{e:phi2}.
 Indeed, in the case of the vacuum
elements the two pairings are canonically identified (see
\eqref{e:pairing=ip} below and Section 5).

Let us deal first with the case \eqref{e:Fockpairing1}. We give
first the invariant definition, and then the `constructive'
definition along the lines of $< \ , \ >_{W}$ in Section 2.
Invariantly, in the case $W_1 = \wt{W}_1$, the pairing
$\k:\DET_{W_1} \times \DET_{W_1\pp}\to \mathbb{C}$ defines an
embedding $\gamma:\DET_{W_1}-0\to \Ff_{W_1\pp}$ by $\gamma(a)( \ .
\ ) = f(a, \ . \ )$, and hence a map $\rho :\Ff_{W\pp}^{*} \to
\Ff_W$, $\rho(f)( \ . \ ) = f(\gamma( \ . \ ))$. This gives us a
pairing $\Ff_{W}^{*}\times \Ff_{W\pp}^{*}\to  \mathbb{C}$ with
$(f,g) = f(\gamma(g))$, and by duality the asserted pairing, since
$\Ff_{W}^{**} \cong \Ff_W$ in the topology of uniform convergence
on compact subsets of $Gr(H)$ ($\psi_{S}\leftrightarrow \ \
evaluation \ \ at \ \ \xi(S)$ cf. \cite{PrSe86} Sect. 10.2,
\cite{Mi85} Sect. 6.2). The general case follows in the same way
with $\mathbb{C}$ replaced by $\Det(\wt{W}_1,W_1)$.

Constructively, recall that  any section in $\Ff_W$ can be written
as a linear combination of the $\psi_{[\a,\la]}$, with
$[\a,\la]\in \DET_W$. Hence for $[\a,\la]\in \DET_{W_1},
[\b,\mu]\in \DET_{\wt{W}_1\pp}$ we can define the Fock pairing by
setting
\begin{equation}\label{e:Fpairing}
  (\psi_{[\a,\la]},\psi_{[\b,\mu]}) =\k
(\psi_{[\a,\la]},\psi_{[\b,\mu]})\in \Det(\wt{W}_1,W_1),
\end{equation}
and then extending by linearity. In particular, from
\eqref{e:detvac} we have $\nu_{W_1} = \psi_{[id_{W_1},1]}$ and
$\nu_{\wt{W}_1\pp} = \psi_{[id_{\wt{W}_1\pp},1]},$ and so
$$(\nu_{W_1},\nu_{\wt{W}_1\pp}) =
\k(det(id_{W_1}),det(id_{\wt{W}_1 \pp}) ) =
det(P_{\wt{W}_1,W_1}),$$

 where the final equality is equation
\eqref{e:detpairing6}. Notice further that if we extend $g_{\phi}$
in \eqref{e:phi2} to a map
  $g_{\phi} : \DET_{W_1} \times \DET_{\wt{W}_1} \to \Det(\wt{W}_1,W_1)$
by  $g_{\phi}([\a,\la],[\b,\mu]) = \ol{\la}\mu
det(\a^{*}P_{W}\b),$ then the Fock space inner-product becomes a
Hermitian pairing $< \ , \ , >_{W_1}
:\Ff_{W_1}\times\Ff_{\wt{W}_1}\to \Det(\wt{W}_1,W_1)$ and with
respect to the identification $Gr(H,W_1)\leftrightarrow
Gr(\ol{H}_1,W_1\pp), \ W\leftrightarrow W\pp$ we have
$\nu_{\wt{W}_1\pp}\leftrightarrow\nu_{\wt{W}_1}$ and

\begin{equation}\label{e:pairing=ip}
(\nu_{W_1},\nu_{\wt{W}_1\pp}) = <\nu_{W_1},\nu_{\wt{W}_1}
>_{W_1} =
det(P_{\wt{W}_1, W_1}).
\end{equation}

The pairing \eqref{e:Fockpairing4} now follows from
\eqref{e:Fockpairing1} and \eqref{e:Pe}. Alternatively we can
define it directly as $ (\psi_{[\a,\la]},\psi_{[\b,\mu]}) =
\psi_{\k ([\a,\la],[\b,\mu])}$, where $\k$ is the pairing
\eqref{e:detpairing1}. Note that if conditions (i) and (ii) do not
hold then $\k ([\a,\la],[\b,\mu]) = 0$. Equation
\eqref{e:Fockpairing5a} is now just by construction, and equation
\eqref{e:Fockpairing5} follows easily from \eqref{e:detpairing8}.
\end{proof}

The Fock space pairing \eqref{e:Fockpairing3} defines an
isomorphism $\Ff_{W_{1}\pp} \cong \Ff_{W_{1}}^{*}$ and hence the
vector  $\phi_{\e}\in \Ff_{W_{1}\pp}(\ol{H}_1)\otimes
\Ff_{W_{2}}(H_2)$ defined by $\e\in
\Mor_{\Cc_{Gr}}((H_1,W_{1}\pp),(H_2,W_2))$ is canonically an
element of $\Hom(\Ff_{W_{1}}(H_1), \Ff_{W_{2}}(H_2))$ which is a
morphism of $\Cc_{vect}$, as required. In the case
$(\ol{H}_0,W_0\pp) = \O$ the map $\Mor_{\Cc_{Gr}}(\O,(H_1,W_1))
\to \Hom(\CC,\Ff_{W_1})$ is defined by $1\mto \nu_{W}$, and
similarly when $(H_2,W_2) = \O$. The functoriality of the
composition of linear maps with respect to the multiplication in
$\Cc_{Gr}$ is precisely \eqref{e:Fockpairing5a}. We may state this
as:
\begin{thm}\label{t:pairing}
The category multiplication in $\Cc_{Gr}$ induces through the Fock
space functor a canonical multiplication in the category
$\Cc_{vect}$.
\end{thm}
This can be conveniently summarized in the statement that the
following diagram commutes:

$$\begin{CD} Gr(\ol{H}_0\oplus H_1, W_0\pp\oplus W_1) \times
Gr(\ol{H}_1\oplus H_2, W_1\pp\oplus W_2) @>>{*}> Gr(\ol{H}_0\oplus
H_2, W_0\pp\oplus W_2)\\
  @VV{(\e,\d)}V @VV{\e*\d}V  \\
 \DET_{W_0\pp\oplus W_1} \times \DET_{W_1\pp\oplus W_2}
@>>{\k}>   \DET_{W_0\pp\oplus W_2} \\  @VV{{\rm Plucker}}V
@VV{{\rm Plucker}}V
\\
\Ff_{W_0\pp}\otimes\Ff_{W_1} \otimes \Ff_{W_1\pp}\otimes\Ff_{W_2}
@>>{( \ , \ )}> \Ff_{W_0\pp}\otimes\Ff_{W_2}
\end{CD},$$

\bigskip

\noi where $\e,\d$ are, respectively,  a choice of section of the
bundles $ \DET_{W_0\pp\oplus W_1}$ and $\DET_{W_1\pp\oplus W_2}$.

\subsection{The Fock Functor}

The Fock functor $Z : \Cc_d \to \Cc_{vect}$ is the projective
functor defined by the  composition of the functors $\Psi$ and
$\Phi$, thus $Z$ is the functor
\begin{equation}\label{e:Ffunctor}
  Z = \Phi \circ \Psi : \wt{\Cc}_d \too \Cc_{vect}.
\end{equation}

\noi $Z$ acts on objects of $\wt{\Cc}_d $ by
\begin{equation}\label{e:Fobj}
  Z  : \Ob(\wt{\Cc}_d) \too \Ob(\Cc_{vect}),
\end{equation}
$$ Z((\Yy,W)) = \Ff_{W}(H_Y), \hskip 5mm Z(\O) = \mathbb{C},$$

\noi and on morphisms by
\begin{equation}\label{e:Fobj2}
  Z  : \Mor_{\wt{\Cc}_d}((\Yy_1,W_1),(\Yy_2,W_2)) \too
  \Mor_{\wt{\Cc}_d}(\Ff_{W_1}(H_{Y_1}) , \Ff_{W_2}(H_{Y_2})),
\end{equation}
$$ Z((\Xx,\e)) = \phi_{\e}, \hskip 5mm \e\in\Det(W_1\pp\oplus W_2,
K(D_X)), $$ where $(\Yy_1,W_1)$,$(\Yy_2,W_2)$ are not both empty,
and $\phi_{\e}$ is defined as in Section 3.6 by the Fock space
pairing. If $(\Yy_1,W_1) = \O =(\Yy_2,W_2)$, so $X$ is a closed
manifold, then $$Z(\Xx) = det(D_X) \in \Det(D_X)\cong
\mathbb{C},$$ where the trivialization requires a choice.

The `sewing property' of the FQFT is precisely the functorial Fock
space pairing of \propref{p:Fockpairing}. Note that if both
$W_i\neq \O$ it is not possible to choose $\phi_{\e}$ to be the
vacuum vector $\nu_{W_1\pp\oplus W_2}\in\Ff_{W_1\pp\oplus W_2}$,
since $K(D_X)$, depending on global data, is always transverse to
the pure boundary data $W_1\pp\oplus W_2$. Consider though the
case $W_1 = \O$. Let $X$ be a closed connected manifold
partitioned by an embedded codimension 1 submanifold $Y$, so that
$X = X^0 \cup_Y X^1$. Here $X^0, X^1$ are manifolds with boundary
$Y$, where $\dd X^0 = \ol{Y}$ has outgoing orientation and $\dd
X^1 = Y$ has incoming orientation. $X^0$ is assumed to be
associated to a morphism $\Xx^0$ in $\Mor_{\Cc_d}(\O,(\Yy,W))$ for
a choice of admissible polarization $W \in Gr(H_Y)$. In this case
we {\em can} choose in particular $W = K(D^0)$ and $\phi_{\e} =
\nu_{K(D^0)}$. Similarly, we have $\Xx^1\in
\Mor_{\Cc_d}((\Yy,W\pp),\O)$,  and we may choose $W\pp = K(D^1)$.
As a Corollary of the properties of $Z$ we then have the following
algebraic sewing law for the determinant with respect to a
partitioned closed manifold.
\begin{thm}\label{t:sewing}
There are functorial bilinear pairings
\begin{equation}\label{e:sewing1}
( \ , \ ) : \Ff_{K(D^0)}(H_Y) \times \Ff_{W\pp}(\ol{H}_Y) \too
\Det(D^{0}_{P_W}),
\end{equation}
where the right-side is the determinant line of the EBVP $D_P$
with
\begin{equation}\label{e:sewing2}
( \nu_{K(D^0)} , \nu_{W\pp} ) = det(D^{0}_{P}),
\end{equation}
and
\begin{equation}\label{e:sewing3}
( \ , \ ) : \Ff_{K(D^0)}(H_Y) \times \Ff_{K(D^1)}(\ol{H}_Y) \too
\Det(D_X),
\end{equation}
where the right-side is the determinant line of the Dirac operator
$D_X$ over the closed manifold $X$, with
\begin{equation}\label{e:sewing4}
( \nu_{K(D^0)} , \nu_{K(D^1)} ) = det(D_X).
\end{equation}
\end{thm}
\begin{proof}
We just need to recall a couple of facts. From equations
\eqref{e:Fockpairing1} and \eqref{e:Fockpairing2}  we have a
pairing $ \Ff_{K(D^0)}(H_Y) \times \Ff_{W\pp}(\ol{H}_Y) \too
 \Det(K(D^0),W)$ with $( \nu_{K(D^0)} , \nu_{W\pp} ) = det(S(P_W))$,
 where $S(P_W):K(D^0)\to W$ is the operator of Section 2. But from
\eqref{e:isom3} there is a canonical isomorphism $\Det(S(P_W))
\cong \Det(D_{P_W})$, with $det(S(P_W))\mto det(D_{P,b})$. This
proves the first statement. The second statement follows similarly
upon recalling from \cite{Sc98} (Theorem 3.2) that there is a
canonical isomorphism $\DET((I-P(D^1)\circ P(D^0)) \cong
\Det(D_{X})$, again preserving the determinant elements.
\end{proof}

Thus one may think of the determinant $det(D_P)$ `classically' as
an object in the complex line $\Det(K(D),P)$ depending on a choice
of boundary condition $P$, or absolutely as a `quantum
determinant' as a ray in the Fock space $\Ff_{K(D)}$ defined by
the vacuum vector that does not depend on a choice of $P$. The two
view points being related by \eqref{e:sewing2}.

\bigskip

Finally, we point out that, in particular, the Fock functor
naturally defines a map from geometric fibrations to vector
bundles. To a geometric fibration $\NN$ of closed $d$-dimensional
manifolds endowed with a spectral section $\Pf$ it assigns the
corresponding Fock bundle $\Ff_{\Pf}$. A `projective' morphism
between objects $(\NN_1,\Pf_1)$ and $(\NN_2,\Pf_2)$ is a geometric
fibration of $\MM$ of $d+1$-dimensional manifolds with boundary
$\NN_1 \sqcup \NN_2$ along with a section of the determinant
bundle $\DET(\Pf_1\pp\oplus\Pf_2, K(\Df))$, where $\Df$ is the
family of Dirac operators defined by $\MM$.  This defines a bundle
map $\Ff_{\Pf_1}\to \Ff_{\Pf_2}$  using the generalized Plucker
embedding \eqref{e:genPlucker} and the Fock space pairing. For a
partition of a closed geometric fibration $M = M^0 \cup_{N}M^1$
over a parameter manifold $B$ by an embedded fibration of
codimension 1 manifolds, the analogue of \thmref{t:sewing} then
states that there are functorial Fock bundle pairings:
\begin{equation}\label{e:bsewing1} ( \ , \ ) : \Ff_{P(\Df^0)}
\times \Ff_{\Pf\pp} \too \DET(\Df^0, \Pf),\hskip 5mm (
\nu_{P(\Df^0)} , \nu_{\Pf\pp} ) = det(D^{0}_{\Pf}),
\end{equation}
where the right-side is the determinant line bundle of the family
of  EBVP $(\Df^0,\Pf)$, $\nu_{\Pf}$ is the vacuum section of the
Fock bundle $\Ff_{\Pf}$ and $det(D^{0}_{\Pf})$ the determinant
section of $\DET(\Df^0, \Pf);$ and
\begin{equation}\label{e:bsewing3}
( \ , \ ) : \Ff_{P(\Df^0)}\times \Ff_{P(\Df^1)} \too
\Det(\Df_{\MM}),\hskip 5mm ( \nu_{K(\Df^0)} , \nu_{K(\Df^1)} ) =
det(D_{\MM}).
\end{equation}

The proof again requires only the properties of the Fock bundle
pairing and the determinant bundle identifications of Section 2
and \cite{Sc98}. Notice that there is no regularization here of
the determinant, but only a pairing between bundle sections.


\section{Gauge Anomalies and the Fock Functor}

In this Section we give a physical application of these ideas with
a Fock functor description of the chiral and commutator anomalies
for an even-dimensional manifold with (odd-dimensional) boundary.

The Fock functor assigns vector spaces to all odd-dimensional
compact oriented  spin manifolds $Y$ and polarizations. There is
no further restriction on the topology of $Y.$ However, in this
section we shall restrict to a fixed topological type for $Y.$ For
our purposes this is no real restriction since our principal aim
is to understand the action of continuous symmetries,
diffeomorphisms and gauge transformations, on the family of Fock
spaces and on the morphisms between the Fock spaces; the action of
the symmetry group cannot change the topological type of $Y.$ In
order to be even more concrete, to begin with, we shall consider
the case of the parameter space $B={\mathcal A}$ of smooth vector
potentials labeling the geometries over $Y.$

Thus we are lead to consider the action of the group of gauge
transformations on the bundle ${\mathcal F}$ of Fock spaces over
the base  ${\mathcal A}.$ The gauge transformations act naturally
on the base $\mathcal A$ and thus we have a lifting problem:
Construct a (projective) action of the gauge group in the total
space of $\mathcal F$ intertwining with the family of quantized
Dirac Hamiltonians in the fibers. We want to stress that we are
not going to construct a representation of the gauge group in a
single Fock space but we have have a linear isometric action
between different fibers of the Fock bundle.

First, we recall some known facts about gauge anomalies in even
dimensions. Let $M$ be a closed even-dimensional Riemannian spin
manifold and let  $\Aa$ be the space of vector potentials on a
trivial complex $G$-bundle over $M$. For each $A\in\Aa$ we have a
coupled Dirac operator $D_{A}:\Ci (M;S\otimes E)\to \Ci
(M;S\otimes E)$ given locally by

$$ D_A = \sum_{i=1}^{n}\sigma_{i}\left(\dd_i + \Gamma_{i} +
A_{i}\right),$$

\noi where $\Gamma_i$ and $A_i$ are respectively the components of
the local spin connection and and $G$-connection $A$, and
$\sigma_i$ the Clifford matrices. Since $M$ is even-dimensional,
then $D_A$ splits into positive and negative chirality components,
and the object of interest is the Chiral Dirac operator

$$D^{+}_{A} = D_{A}(\frac{1+\gamma_{n+1}}{2}):\Ci (M;S^{+}\otimes
E)\to \Ci (M;S^{-}\otimes E).$$

Acting on $\Aa$ we have the group of based gauge transformations
$\Gg$, which acts covariantly  on the Dirac operators $D^{+}_{g.A}
= g^{-1} D^{+}_{A} g$, so that $\Ker D_{g.A}^{\pm} = g(\Ker
D_{A}^{\pm})$. We are interested in the Fermionic path integral:
\begin{equation}\label{e:fermidet}
Z(A) = \int_{\Ci (M;S^{+}\otimes E)}
e^{\int_{M}\psi^{*}D^{+}_{A}\psi\,dm}\Dd\psi\Dd\psi^{*},
\end{equation}
and a formal extension of finite-dimensional functional calculus
gives $$Z(A) := \det (D_{A}^{+}).$$ To obtain an unambiguous
regularization of \eqref{e:fermidet} we therefore require a gauge
covariant regularized determinant varying smoothly with $A$ in
order that $Z(A)$ pushes down smoothly to the moduli space $\Mm =
\Aa/\Gg$. In the case of Dirac fermions (both chirality sectors)
this can be done and there is a gauge invariant regularized
determinant det$_{reg}(D_A).$ For chiral Fermions on the other
hand, there is an obstruction due to the presence of zero modes of
the Dirac operator. The covariance of the kernels means that the
determinant line bundle descends to $\Mm$ and the obstruction to
the existence of a covariant $Z(A)$ varying smoothly with
$A\in\Aa$ is the first Chern class of the determinant bundle on
$\Mm$, which is the topological chiral anomaly. A 2-form
representative for the Chern class $\DET\Df^{+}$ can be
constructed as the transgression of the 1-form $\omega_1\in\Omega
(\Gg)$

$$ \omega_1(g) = \frac{1}{2\pi i} \frac{d (\det_{r}(D_{gA}^{+})}
{(\det_{r}(D_{gA}^{+})},$$

\noi  measuring the obstruction to gauge covariance of a choice of
regularized determinant $\det_{r}(D_{A}^{+})$. For details see
\cite{AtSi84,Mi89}.

In the case of a manifold $X$ with boundary $Y$ new complications
arise. Fixing an elliptic boundary condition (spectral section)
$\Pf$ for the family of chiral Dirac operators $\Df^+ = \{
D_{A}^{+}:A\in\Aa\}$, we obtain a Fock bundle $\Ff_{\Pf}$ over
$\Aa$ to which we aim to lift the $\Gg$ action. It is natural to
look first at gauge transformations (or diffeomorphisms) which are
trivial on the boundary. In fact, the calculation of the Chern
class  in \cite{AtSi84} can be extended to this case using a
version of the families index theory for a manifold with boundary,
\cite{BiFr86, P96}. The gauge variation of the chiral determinant
can be written as
\begin{equation} \label{e:anom}
{\det}_r (D_{g.A}^+) = {\det}_r(D_A^+) \omega(g;A)
\end{equation}
where $\log\omega$ is an integral over $X$ of a local differential
polynomial in $g$, $A$ and the metric on $X;$ $\omega$ is the
integrated version of the `infinitesimal' anomaly form $\omega_1.$
The important point is that the formula applies both to the case
of a manifold with/without boundary. In fact, in the latter case
this gives a direct way to define the determinant bundle over
$\Aa/\Gg,$ \cite{Mi87}.

The locality of the anomaly \eqref{e:anom} is compatible with the
formal sewing formula \eqref{e:pisewing}. Applying a gauge
transformation which is trivial on $Y$ to the right-hand-side of
the equation gives a gauge variation which is a product of gauge
variations on the two halves $X_0, X_1$ of $M.$ This product is
equal, by locality of the logarithm, to the gauge variation on $M$
of the path integral on the left-hand-side. Since the cutting
surface $Y$ is arbitrary, one can drop the requirement that $g$ is
trivial on $Y.$

The gauge transformations (and diffeomorphisms) which are not
trivial on the boundary need a different treatment. This is
because they act non-trivially on the boundary Fock spaces
$\mathcal{F}_Y.$ We shall concentrate on the case when $Y$ is odd
dimensional. The first question to ask is how the action of the
gauge group on the parameter space $B$ of boundary geometries on
$Y$ is lifted to the total space of the bundle of Fock spaces
${\mathcal F} \to B.$ This problem has already been analyzed
(leading to Schwinger terms in the Lie algebra of the group $\Gg$)
in the literature, but in the present article we want to clarify
how the boundary action intertwines with the Fock functor
construction.

\subsection{Commutator Anomaly on the Boundary}

Let $b\in B$ and $W\in Gr_b.$ In the rest of this section $B$
denotes the space of metrics and vector potentials on a fixed
manifold $Y$ and $Y_b$ is the manifold $Y$ equipped with the
geometric data $b.$ The pair $(b,W)$ is mapped to $(g.b, g.W)$ by
a gauge transformation (or a diffeomorphism) $g,$ acting on both
potentials, metrics and spinor fields. This induces an unitary map
from the Fock space $\mathcal F(H_b,W)$ to $\mathcal
F(H_{g.b},g.W),$ by $a^*(u) \mapsto a^*(g\cdot u)$ and similarly
for the annihilation operators.

However, sometimes $(b,W)$ do not appear independently, but $W$ is
given as a function of the boundary geometry; $b\to P_{W=W_b}$ is
a Grassmann section; this leads to the construction of the bundle
of Fock spaces ${\mathcal F}_b$ parameterized by $b\in  B,$ as
already mentioned above. An example of this situation is the
following. Suppose the Dirac operators on the boundary do not have
zero eigenvalue (this happens when massive Fermions are coupled to
vector potentials). Then it is natural to take $W_b = H^{+}_{b}$
as the space of positive energy states. Still this case does not
lead to any complications because of the equivariance property
$W_{g.b} = g.W_b.$ However, there are cases when no equivariant
choices for $W_b$ exist. This happens when we have massless chiral
Fermions coupled to gauge potentials. For some potentials there
are always zero modes and one cannot take $W_b$ as the positive
energy subspace without introducing discontinuities into the
construction.

Let us assume that a Grassmann section $W_b$ is given. For each
boundary geometry $b$ we have a Fermionic Fock space ${\mathcal
F}_b=\mathcal{F}(H_b,W_b)$ determined by the polarization
$H_{Y_b}=W_b\oplus W_b^{\perp}.$ In order to determine the
obstruction to lifting the gauge group action on $Y$ to the bundle
of Fock spaces such that $g^{-1} D_{Y_b} g= D_{Y_{g.b}}$ we
compare the action on $\mathcal F$ to the natural action in the
case of  polarizations $W'_b$ defined by the positive energy
subspaces of Dirac operators $D_{Y_b} -\lambda.$ We have fixed a
real parameter $\lambda$ and we consider only those boundary
geometries $b\in B$ for which $\lambda$ is not an eigenvalue.
Since the choice of polarizations $W'$ is equivariant, the gauge
action lifts to the (local) Fock bundle $\mathcal F'.$ Relative to
$W$ the $\mathcal F'$ vacua form a complex line bundle
$\DET(W',W);$ again, this is defined only locally in the parameter
space.\\[1mm]

\noi {\bf Example 4.1}\;\; Let $Y$ be a unit circle with standard
metric but varying gauge potentials. We can choose $H_Y=W\oplus
W^{\perp}$ as the fixed polarization defined by the decomposition
to positive and negative Fourier modes. If the gauge group is
$SU(n)$ and Fermions are in the fundamental representation of
$SU(n)$ then the mapping $g\mapsto g\cdot W$ defines an embedding
of the loop group $LSU(n)$ to the Hilbert-Schmidt Grassmannian
$Gr_1(H_Y,W).$  The pull-back of the Quillen determinant bundle
over the Grassmannian to $LSU(n)$ defines the central extension of
the loop group with level $k=1,$ \cite{PrSe86}.\\[1mm]

There is a general method to describe the relative determinant
bundle in terms of index theory on $X$ for $\partial X=Y.$ We
assume that the spin and gauge vector bundle on $Y$ can be
smoothly continued to bundles on $X.$ This is the case for example
when $X=S^{2n-1}$ and $Y$ is chosen such that it has the topology
of a solid ball, with a product metric near the boundary. Any
vector potential can be smoothly continued to a potential on $X$
for example as $A(x,r)= f(r) A(x)$ with $f$ increasing smoothly
from zero to the value one at $r=1;$ all derivatives of $f$
vanishing at $r=0,1.$ We can now define a spectral section
$b\mapsto W_b$ as the Calderon subspace associated to the
continued metric and vector potential in the bulk; we denote the
Dirac operator defined by this geometric data in $X$ by $D_{X,b}.$
The determinant line for a Dirac operator $D_{X,b}$ subject to the
boundary condition $W$  is canonically the tensor product of the
line $\DET(W',W)$ and the determinant line of the same operator
$D_{X,b}$ but subject to another choice of boundary conditions
$W',$ \eqref{e:isom1}

Since the spectral section $W_b$ and the Dirac operator $D_{X,b}$
is parameterized by the affine space of geometric data (metrics
and potentials)  on the boundary, the corresponding Dirac
determinant bundle is topologically trivial. Let $U_{\lambda}$ be
the set of $b\in B$ such that the real number $\lambda$ is not in
the spectrum of the corresponding Dirac operator $D_{Y_b}.$ On
$U_{\lambda}$ we can define the boundary conditions
$W'_{z,\lambda}$ as the spectral subspace $D_{Y_b} >\lambda$ of
the boundary Dirac operator. The set $U_{\lambda}$ is in general
non-contractible and the Dirac determinant line bundle defined by
the boundary conditions $W'$ can be nontrivial.  The curvature of
this bundle is given by the families index theorem \cite{BiFr86,
P96}. It can be written in terms of characteristic classes in the
bulk and the so-called $\eta$-form on the boundary; the latter
depends on spectral information about the family of Dirac
operators. The curvature $\Omega$ when evaluated along gauge and
diffeomorphism directions on the boundary data has a simplified
expression; in particular, the $\eta$-form drops out since it is a
spectral invariant and the contribution from the characteristic
classes in the bulk reduces to a boundary integral involving the
(gauge and metric) Chern-Simons forms, \cite{CMM97, EM99}:
\begin{equation}\label{e:curv}
\frac{1}{2\pi}\Omega  =  \int_Y  CS_{[2]}(A+v, \Gamma + w)
\end{equation}
with $$ d CS(A,\Gamma) = \hat A (R) \text{ch} (F) $$
 where $\left[ 2\right]$ denotes the part that is a 2-form along parameter
 directions.  The symbol $A+v$ means a connection form on $Y\times B$ such
 that in the $Y$ directions it is given by a vector potential $A$
 and in the gauge directions ${\mathcal L}_u$ on $B$ it is equal to the Lie
 algebra valued function $u.$ In a similar way, $\Gamma + w$ is the sum
 of the Levi-Civita connection (on $Y$) and a metric connection $w$ such
 that the value of $w$ along a vector field ${\mathcal L}_u$ on $B$, generated
 by a vector field $u$ on $Y$, is equal to the matrix valued function on $Y$
 given by the Jacobian of the vector field $u.$

The characteristic classes are
\begin{eqnarray*}
\hat A (X) &=& \text{det}^{1/2} \left( \frac{i R /4\pi}{\sinh( i
R/4\pi)}\right)  \\ \text{ch}(X) &=& \tr (\exp(i F/2\pi))
\end{eqnarray*}
where $R$ is the Riemann curvature tensor associated to a metric
$g$ in the bulk, $F$ is the curvature of a gauge connection $A.$

>From the previous discussion it follows that the topological
information (de Rham cohomology class of the curvature) in the
relative determinant bundle $\DET(W',W)$ is given by the curvature
formula for the Dirac determinant bundle for boundary polarization
$W'.$ This leads to the explicit formula for lifting the gauge and
diffeomorphism group action from the base $B={\mathcal
A}\times\mathcal M$  to the Fock bundle $\mathcal F,$
\cite{CMM97}, \cite{EM99}. Infinitesimally, the lifting leads to
an extension of the Lie algebras $Lie(\mathcal G)$ and $Vect(Y)$
by an abelian ideal $J$ consisting of complex valued functions on
${\mathcal A} \times {\mathcal M}.$ The commutator of two pairs of
elements $(u, f)$ and $(v,g)$ (where $f,g$ are in the extension
part $J$ and $u,v$ are infinitesimal gauge transformations or
vector fields) is given as
\begin{equation}
[(u,f), (v,g)] = ([u,v], {\mathcal L}_u\cdot g -{\mathcal
L}_v\cdot f +c(u,v))
\end{equation}
where $c(u,v)$ is an anti-symmetric bilinear function of the
arguments $u,v$ taking values in the ideal $J.$ It satisfies the
cocycle condition
\begin{equation}
c(u_1, [u_2, u_3]) + {\mathcal L}_{u_1} \cdot c(u_2,u_3) +\text{
cyclic permutations } =0.
\end{equation}
The cocycle $c$ is just the curvature form evaluated along gauge
 (or diffeomorphism group) directions,
$$c(u,v)= \Omega(\mathcal{L}_u, \mathcal{L}_v)$$ where
$\mathcal{L}_u$ is the vector field on $\mathcal A$ (resp.
$\mathcal M$) generated by the gauge (diffeomorphism) group
action. When $Y$ is one-dimensional, the cocycle reduces to the
central term in an affine Lie algebra or in the Virasoro algebra;
in this case the cocycle does not depend on the vector potential
or the metric on $Y.$   In dimension 3 the cocycle (Schwinger
term) is given as \cite{Mi85, FSh}
\begin{equation}
c(u,v)= \frac{i}{24\pi^2} \int_Y \tr\, A [du, dv]
\end{equation}
when the Fermions are in the fundamental representation of the
gauge group; here $u,v: Y\to G$ are smooth infinitesimal gauge
transformations. In dimension 3 the cocycle is trivial in case of
vector  fields and metrics. Also in higher dimensions explicit
expressions can be worked out starting from \eqref{e:curv},
\cite{EM99}.

The curvature of the relative determinant bundle, in the case of
Grassmann sections $W,W'$ discussed above, can be written as
\begin{equation} \label{e:curv2}
\omega(u,v)= \text{tr}\,(F'_b\mathcal{L}_u F'_b\mathcal{L}_v F'_b
-F_b\mathcal{L}_u F_b\mathcal{L}_v F_b),
\end{equation}
where $F_b= P_b -P_b^{\perp}$, $P_b$ is the orthogonal projection
onto $W_b$ and $P_b^{\perp}$ is the projection on to the
orthogonal complement. Note that neither of the two terms on the
right have a finite trace  but the difference is trace class by
the relative trace-class property of $P_b,P'_b.$ Note also that in
the case when all the projections $P'$ are in a single restricted
Grassmannian, the first term is the standard formula for the
curvature of the Grassmannian. The second term can be viewed as a
renormalization; it is in fact a background field dependent vacuum
energy subtraction.

The proof of the curvature formula \eqref{e:curv2}  is as follows.
First, one notices that this gives the curvature of the relative
determinant bundle when both variables $W_b,W'_b$ lie in the same
restricted Grassmannian relative to a fixed base point $P_0.$ Then
one has to show that the difference actually makes sense when
dropping the existence of common base point. For that purpose one
writes $$\omega(u,v)= \tr\, \left[(F'_b-F_b)\Ll_u  F'_b \Ll_v F'_b
+F_b(\Ll_u F'_b- \Ll_u F_b)\Ll_v F'_b + F_b\Ll_u F_b (\Ll_v
F'_b-\Ll_v F_b)\right],$$ which is manifestly a trace of a sum of
trace-class operators.

\subsection{Chiral Anomaly in the Bulk}

In the construction of the Fock functor we took as independent
parameter a choice of an element $\lambda\in \DET(K(D_b^+),
W_{Y_b})$ in the boundary determinant bundle; recall that
$K(D_b^+)$ is the range of the Calderon projection.  A choice of
this element, as a function of the geometric data in the bulk, is
a section of the determinant bundle. In quantum field theory such
a choice is provided by a choice of the regularized determinant of
the chiral Dirac operator $D_b^+.$ The determinant vanishes if and
only if the orthogonal projection $\pi:K(D_b^+) \to W_{Y_b}$ is
singular and therefore it makes sense to choose $\lambda=\lambda_b
\in \DET(K(D_b^+),W_{Y_b})$ (represented as an admissible linear
map $\lambda_b: W_{Y_b} \to K(D_b^+)$) such that $det_r(D_b^+),$
defined subject to the boundary conditions $W_b,$ is equal to
$\det_F(\pi\circ \lambda_b).$

In the case of chiral Fermions the determinant $det_r(D_b^+)$ is
anomalous with respect to diffeomorphisms and gauge
transformations on $X$ and the variation of the determinant is
given by the factor $\omega(g;b)$ in \eqref{e:anom}. This implies
the transformation rule
\begin{equation}\label{e:anomaly2}
\lambda_{g.b}= \lambda_b \cdot \omega(g;b)
\end{equation}
where $g$ is either a gauge transformation or a diffeomorphism and
$b$ stands for both the metric and gauge potential on $X.$
$\omega$ is a non-vanishing complex function, satisfying the
cocycle condition
\begin{equation} \label{e:cocyc}
\omega(g_1 g_2;b) =\omega(g_1;g_2\cdot b) \omega(g_2;b).
\end{equation}
Here the boundary conditions should be invariant under $g,$
meaning that the gauge transformations (and diffeomorphism)
approach smoothly the identity at the boundary.

If the cocycle $\omega$ is nontrivial (and this is the generic
case for chiral Fermions) in cohomology, then  the relation
\eqref{e:anom} above tells us that the Fock functor is determined
by the family of Calderon subspaces $K(D_b^+)$ and a choice of a
section (the regularized determinant) of a nontrivial line bundle
over the quotient space $\mathcal B$ of $B$ modulo diffeomorphisms
and gauge  transformations. \\[1mm]

\noi {\bf Example 4.2}\;\;  Let ${\mathcal A}(D)$ be the space of
smooth potentials in a unit disk $D.$ Let ${\mathcal G}(D,\partial
D)$ be the group of gauge transformations which are trivial on the
boundary $\partial D= S^1.$ For each $A\in {\mathcal A}(D)$ there
is a unique $g=g_A:D\to G$ such that $A'= g^{-1} Ag + g^{-1}dg$ is
in the radial gauge, $A_r=0,$ and $g(p)=1,$ where $p\in S^1$ is a
fixed point on the boundary. It follows that ${\mathcal
B}={\mathcal A}(D)/{\mathcal G} (D,\partial D)$ can be identified
as ${\mathcal A}_{rad}(D) \times {\mathcal G}(D)/{\mathcal
G}(D,\partial D) = {\mathcal A}_{rad}\times \Omega G,$ where
${\mathcal A}_{rad}(D)$ is the set of potentials in the radial
gauge and $\Omega G$ is the group of based loops, i.e., those
loops in $G$ which take the value $1$ at the point $p.$ The first
factor ${\mathcal A}_{rad}(D)$ is topologically trivial as a
vector space. Thus in this case the topology of the Dirac
determinant bundle over  the moduli space $\mathcal B$ is given by
the pull-back of the canonical line bundle  over $\Omega G$,
\cite{PrSe86}, with respect to the map $A \mapsto g_A|_{\partial
D}.$  The  sections of $\DET \to \mathcal B$ are by definition
complex functions $\lambda : {\mathcal A}(D) \to \Bbb C$ which
obey the anomaly condition \eqref{e:anom}, \cite{Mi87}.
\\[2mm]

\subsection{Relation of the Chiral Anomaly to the Commutator Anomaly}

The bulk anomaly and the extension (Schwinger terms) of the gauge
group on the boundary are closely related, \cite{Mi87}. As we saw
above, the Fock functor is determined by a choice of a section
$b\to \lambda_b$ of the relative line bundle
$\DET(K(D_b^+),W_{Y_b}).$  The section transforms according to the
chiral transformation law for regularized determinants,
$$\lambda_{g\cdot b}  = \omega(g;b) \lambda_b,$$ for
transformations $g$ which are equal to the identity on the
boundary. If now $h$ is a transformation which is not equal to the
identity on the boundary, we can define an operator $T(h)$ acting
on sections by
\begin{equation} \label{e:oper}
(T(h) \xi )(b)= \gamma(h;b) \xi(h^{-1}\cdot b),
\end{equation}
 where $\gamma$ is a complex function of modulus
one and must be chosen in such a way that $\xi'(b)= (T(h)\xi)(b)$
satisfies the condition \eqref{e:anom}.  Explicit expressions for
$\gamma$ have been worked out in several cases, \cite{Mi89}. For
example, if dim$X=2$ and $g$ is a gauge transformation then
\begin{equation} \label{e:gamma}
\gamma(h;A) = \exp(\frac{i}{2\pi}\int_X \tr A\, dh h^{-1}),
\end{equation}
where $\tr$ is the trace in the representation of the gauge group
determined by the action on Fermions.  In general, $\gamma$ must
satisfy the consistency condition,
\begin{equation} \label{e:cons}
\gamma(h;g\cdot b) \omega(hgh^{-1}; h^{-1}\cdot b)=
\gamma(h;b)\omega(g;b).
\end{equation}
 In the two-dimensional gauge
theory example,
\begin{equation} \label{e:omega}
\omega(g;A)= \exp(\frac{i}{2\pi}\int_X  \tr A dg g^{-1} +
\frac{i}{24\pi} \int \tr (dg g^{-1})^3).
\end{equation}
The latter integral is evaluated over a 3-manifold $M$ such that
its boundary is the closed 2-manifold obtained from $X$ by
shrinking all its boundary components to a point.

The introduction of the factor $\gamma$ in \eqref{e:oper} has the
consequence that the composition law for the group elements is
modified,
\begin{equation} \label{e:theta}
T(g_1) T(g_2) = \theta(g_1, g_2;z) T(g_1 g_2),
\end{equation}
where $\theta$ is a $S^1$ valued function, defined by
\begin{equation} \label{e:theta2}
\theta(g_1,g_2;b) = \gamma(g_1g_2;b) \gamma(g_1;b)^{-1}
\gamma(g_2; {g_1}^{-1}\cdot b).
\end{equation}
Thus we have extended the original group of gauge transformations
(diffeomorphisms) by the abelian group of circle valued functions
on the parameter space $B.$

At the Lie algebra level, the relation \eqref{e:theta} leads to a
modified commutator (by Schwinger terms discussed above) of the
`naive' commutation  relations of the algebra of infinitesimal
gauge transformations or the algebra of vector fields on the
manifold $X.$ Actually, the modification is `sitting on the
boundary'; the action of $T(g)$ was defined in such a way that the
(normal) subgroup of gauge transformations which are equal to the
identity on the boundary acts trivially on the sections $\xi(b).$
There is an additional slight twist to this statement. Actually,
the normal subgroup is embedded in the extended group as the set
of pairs $(g, c(g)),$ where $c(g)$ is the circle valued function
defined by
\begin{equation} \label{e:cg}
c(g)=  \gamma(g;b)^{-1} \omega(g;b).
\end{equation}
The consistency condition \eqref{e:cons}  guarantees that the
multiplication  rule $$(g_1, c(g_1))(g_2,c(g_2))=(g_1g_2,
c(g_1g_2)),$$ holds in the extended group with the multiplication
law $$(g_1,\mu_1)(g_2,\mu_2) = (g_1 g_2, \theta(g_1,g_2)
\mu_1{\mu_2}^{g_1}),$$ where $(\mu^g(b)= \mu(g^{-1}b).$

\subsection{Summary}

 Let us summarize the above discussion on Fock
functors and group extensions. On the boundary manifold
$Y=\partial X$ a choice of boundary conditions $W_b$ (labeled by a
parameter space $B_Y$ of boundary geometries) defines a fermionic
Fock space ${\mathcal F}_{Y_b}.$ The group of gauge
transformations (or diffeomorphisms) on $Y$ acts in the bundle of
Fock spaces (parameterized by geometric data on the boundary)
through an abelian extension; the Lie algebra of the extension is
determined by a 2-cocycle (Schwinger terms) which are computed via
index theory from the curvature of the relative determinant
bundles \DET$(W'_b, W_b)$ where $W'_b$ is the positive energy
subspace defined by the boundary Dirac operator. If the boundary
is written as a union $Y= Y_{in}\cup Y_{out}$ of the ingoing and
outgoing components then the Fock functor assigns to the geometric
data on $X$ a linear operator $Z_X: \mathcal F_{in} \to \mathcal
F_{out}.$ A gauge transformation in the bulk $X$ sends $Z_X$ to
$\gamma(g;X) Z_{g^{-1}X}.$ This action defines an abelian
extension of the gauge group. There is a normal subgroup
isomorphic to the group of gauge transformations which are equal
to the identity on the boundary. This subgroup acts trivially,
therefore giving an action of (the abelian extension of) the
quotient group on the boundary. The latter group is isomorphic to
the group acting in the Fock bundle over boundary geometries.

\bigskip

\section{Path integral formulae and a 0+1-dimensional example}

In this section we outline the fermionic path integral formalism
for an EBVP and explain how the  Fock functor models this
algebraically.

\subsection{Path integral formulae}
The analogue of \eqref{e:fermidet} for an EBVP is
\begin{equation}\label{e:Pfermidet}
Z_{X}(P) := \det (D_P) = \int_{\Ee_{P}}
e^{\int_{X}\psi^{*}D\psi\,dm}\Dd\psi\Dd\psi^{*},
\end{equation}
where $\Ee_P = dom (D_P)$. This is equation \eqref{e:pi} for the
case $S(\psi) = \int_{X}\psi^{*}D\psi\,dx$ and where the local
boundary condition $f$ has been replaced by the global boundary
condition $P$. If we consider a partition of the closed manifold
$M=X_{0}\cup_{Y}X_{1}$. The Dirac operator over $M$ restricts to
Dirac operators $D^{0}$ over $X^{0}$ and $D^{1}$ over $X^1$. We
assume that the geometry is tubular in a neighbourhood of the
splitting manifold $Y$, then we have Grassmannians $Gr_{Y^i}$ of
boundary conditions associated to $D^i$, where $Y^1 = Y =
\ol{Y^{0}}$. The reversal of orientation means that there is a
diffeomorphism $Gr_{Y^0} \equiv Gr_{Y^1}$ given by
$P\leftrightarrow I-P$, so that each $P\in Gr_{Y^0}$ defines the
boundary value problems $D^{0}_{P}$ and $D^{1}_{I-P}$. According
to \eqref{e:fermidet} and \eqref{e:Pfermidet}, the analogue of the
sewing formula \eqref{e:pisewing} is

\begin{eqnarray}\label{e:pisewing}
\int_{\Ee(M)}
e^{\int_{M}\psi^{*}D^{+}_{A}\psi\,dm}\Dd\psi\Dd\psi^{*} & = &
\int_{Gr_Y}\Dd P \left\{ \int_{\Ee_{P}(X^{0})}
e^{\int_{X^{0}}\psi_{0}^{*}D^{0}\psi_{0}\,dx_{0}}\Dd\psi_{0}\Dd\psi_{0}^{*}\right.
\nonumber \\
 &
 \times
 &
 \left.\int_{\Ee_{I-P}(X^{1})}
e^{\int_{X^{1}}\psi_{1}^{*}D^{1}\psi_{1}\,dx_{1}}\Dd\psi_{1}\Dd\psi_{1}^{*}\right\}.
\end{eqnarray}

\noi That is,

\begin{equation}\label{e:detsewing0}
  Z_{M} = \int_{Gr_Y}Z_{X^{0}}(P)Z_{X^{1}}(I-P)\;\Dd P
\end{equation}

\noi or

 \begin{equation}\label{e:detsewing}
  det (D)  = \int_{Gr_Y} det D^{0}_{P}.det D^{1}_{I-P}\;\Dd P.
\end{equation}

\bigskip
Of course, the above formulas are only heuristic extensions to
infinite-dimensions of a well-defined finite-rank linear
functional. According to the properties of the Fock functor (see
Section 3), the Fermionic integral may be rigourously understood
as a linear functional $\wedge (H_Y\oplus \ol{H}_Y) \too \Det
(D^{0}_{P})$, while \eqref{e:detsewing} is replaced by the
evaluation of the Fock space bilinear pairing on vacuum elements
\eqref{e:sewing3}:
\begin{equation}\label{e:fockD}
det (D) = (\nu_{K(D^0)},\nu_{K(D^1)}).
\end{equation}
However, adopting a slightly different point of view gives a more
precise meaning to the integral formulae above. With a given
boundary condition $P$ the determinants of the (chiral) Dirac
operators on the manifolds $X_0$ and $X_1$ should be interpreted
as elements of the determinant line bundle $\DET$ over the
Grassmannian $Gr_Y,$ with base point $H^+.$ The actual numerical
value of the Dirac determinant depends on the choice of a (local)
trivialization. For example, one could define $det(D)$ as the zeta
function regularized determinant $det_{\zeta}((D_B)^* D_A),$ where
$D_B$ is a background Dirac operator chosen in such a way that
$(D_B)^* D_A$ has  a spectral cut, i.e., there is a cone in the
complex plane with vertex at the origin and no eigenvalues of the
operator lie inside of the cone. The value of the zeta determinant
will depend on the choice of the background field $B.$

A choice of an element in the line in $\DET$ over $P^0\in Gr_Y$ is
given by a choice of a pair $(\alpha,\lambda),$ where $\alpha:H_+
\to P^0$ is a unitary map and $\lambda\in \Bbb C.$ It can be
viewed as a holomorphic section of the dual determinant bundle
$\DET^*$ according to \eqref{e:eval},
$$\psi_{[\alpha,\lambda]}(\xi) = \lambda det_F(\alpha^* \circ \pi
\circ\xi),$$ where $\pi : \xi(H_+) \to P^0$ is the orthogonal
projection. We can think of the variable $\xi$ as the parameter
for different elements $W=\xi(H_+) \in Gr_Y.$  We want to replace
the (ill-defined) integral $\int_{Gr_Y}  det(D_P^0) det(D_P^1) dP$
by a (so far ill-defined) integral of the form
\begin{equation} \label{e:intxi}
\int_{\xi} \psi_{[\alpha,\lambda]}(\xi)^* \psi_{[\beta,\mu]} (\xi)
d\xi.
\end{equation}
 But
this integral looks like the functional integral defining the
inner product between a pair of fermionic wave functions (vectors
in the Fock space) defined in equation \eqref{e:sumxi}:
\begin{equation} \label{e:sumxi2}
det_{F}(\a^{*}\b) = <\psi_{[\alpha,\lambda]}, \psi_{[\beta,\mu]}>
= \sum_{S\in \Ss} \psi_{\alpha, \lambda}(\xi(S))^*
\psi_{[\beta,\mu]} (\xi(S)).
\end{equation}
The relation with \eqref{e:fockD} is given by the identity
\eqref{e:pairing=ip} which tells us that
\begin{equation}\label{e:pairing=ip2}
(\nu_{K(D^0)},\nu_{K(D^1)}) = <\nu_{K(D^0)},\nu_{K(D^1)\pp}>,
\end{equation}
so here $[\a,\la] = det(P_{K^0}P_{K^0}), [\b,\mu] = det(P_{K^1}\pp
P_{K^1}\pp)$. To illustrate this consider the case of the Dirac
operator over an odd-dimensional spin manifold $M$ partitioned by
$Y$. In this case we know from \cite{Sc95} that $K(D^0) =
graph(h_0: F^+\to F^-)$ and  $K(D^1) = graph(h_1: F^-\to F^+)$,
where $F^{\pm}$ denotes the spaces of positive and negative spinor
fields over the even-dimensional boundary $Y$, and $h_0$ is a
unitary isomorphism differing from $g_+ =
(D_{Y}^{-}D_{Y}^{+})^{-1/2}D_{Y}^{+}$ by a smoothing operator.
Here $D_{Y}^{\pm}$ are the boundary chiral Dirac operators which
we assume to be invertible. In particular, $H^+ = graph(g_+:
F^+\to F^-)$. Similarly for $h_{1}$ ,with $g_+ =
(D_{Y}^{+}D_{Y}^{-})^{-1/2}D_{Y}^{-}$. The graph description gives
us a canonical trivialization of the determinant lines, so that
$$P(D^0) = \frac{1}{2}\begin{pmatrix}
  Id_{F^{+}} & h_{0}^{-1} \\
  h_0 & Id_{F^{+}}
\end{pmatrix}.$$
Then by the definition \eqref{e:detpairing5} of the Fock pairing
we have in this case, with respect to the trivialization,
\begin{equation}\label{e:pairing=ip3}
(\nu_{K(D^0)},\nu_{K(D^1)}) = det_F (\frac{1}{2}(P(D^0)\oplus
(I-P(D^1):K(D^0)\oplus K(D^1)\pp \too H_{Y})
\end{equation}
where the $1/2$ arises (as can be shown canonically) because
$F^{+}$ is not quite an element of the Grassmannian. This is the
operator
\begin{eqnarray*}(P(D^0)\oplus (I-P(D^1))((s^{+},h_{0}s^{+}), (-h_1
s^{-},s^{-}))
  & =&  ((s^{+} - h_1 s^{-}, h_0 s^{-} + s^{-})) \\
  & = &\begin{pmatrix}
  Id_{F^{+}} & h_{1} \\
  h_0 & Id_{F^{-}}\end{pmatrix}
  \begin{pmatrix}
    s^+ \\
    s^- \\
  \end{pmatrix}.
\end{eqnarray*} So in the graph trivialization
$$(\nu_{K(D^0)},\nu_{K(D^1)}) = det_F \frac{1}{2}\begin{pmatrix}
  Id_{F^{+}} & h_{1} \\
  h_0 & Id_{F^{-}}
\end{pmatrix} = det_{F}(\frac{1}{2}(I - h_1h_2)),$$
using the formula $det_F\begin{pmatrix}
  a & b \\
  c & d
\end{pmatrix} = det(d)det(a-bd^{-1}c),$
valid for matrices of the form $Id + trace-class$ provided
$d:F^-\to F^-$ is invertible.

On the other hand, using \eqref{e:sumxi2}, in the trivialization
given: $$<\nu_{K(D^0)},\nu_{K(D^1)\pp}> =
det_{F}\frac{1}{2}(\alpha_{-h_{1}^{-1}}^{*}\alpha_{h_0}) =
det_{F}(\frac{1}{2}(I - h_1h_2)),$$ where $\alpha_{T}
=
\begin{pmatrix}
  \alpha_{+} \\
  T\alpha_{+}
\end{pmatrix}$
where the column index labels the different vectors of the
canonical basis for the the graph of $T:F^{+}\to F^-$, and the row
labels of  $\alpha_{+}$ label the different coordinates of a
suitable basis for $F^+$. The complex number
$det_{F}(\frac{1}{2}(I - h_1h_2))$ is the so called canonical
regularization of $D_M$ relative to the partition $Y$ (see
\cite{Sc95}). There is a corresponding trivialization for
self-adjoint EBVP and its relation with the $\z$-determinant
regularization is explained in \cite{ScWo98}.

\subsection{A (0+1)-dimensional example}
The motivation for replacing the integration formula
\eqref{e:intxi} by the sum in \eqref{e:sumxi} comes from finite
dimensions. If $H=H_-\oplus H_+$ is a decomposition of a $2N$
dimensional vector space into a pair of orthogonal $N$ dimensional
subspaces then the maps $\alpha,\beta,\xi$ above become (with
respect to the basis $\{e_i\}$ with $i= \pm 1, \pm 2, \dots \pm
N$) $2N \times N$ matrices and we have the matrix identity
$$det(\alpha^* \beta) = \sum_{(i)} det(\alpha^* \xi(i))
det(\xi(i)^* \beta),$$ the sum being over all sequences $-N \leq
i_1<i_2 \dots i_N \leq N$ (with $i_{\nu} \neq 0).$ On the other
hand, it follows from eq. (3.48) in \cite{FKS96}, that the
following integration formula holds in this situation:
\begin{equation}\label{e:1integral}
 det(\alpha^* \beta) = a_N \int d\xi d\xi^* det(\alpha^*\xi)
det(\xi^*\beta) \cdot det(\xi^* \xi)^{-2N-1},
\end{equation}
where $a_N$ is a numerical factor and the last factor under the
integral sign can be incorporated to the definition of the
integration measure. If we consider the basis elements
$\alpha_{-h_{1}^{-1}}, \beta_{h_{0}}$ for linear maps
$h_{i}:H^{+}\to H^{-}$ and integrate over the dense subspace
$U_{graph}$ parameterizing the elements $\xi_{T} =
(\xi^{+},T\xi^{-})$ with $T\in \Hom(H^+,H^- )$, the integral
\eqref{e:1integral} becomes
\begin{equation}\label{e:2integral}
 det(1 - h_{2}h_{1}) = a_N \int dT dT^* det_{H^{+}}(1-h_{2}T)
det_{H^{-}}(1+T^{*}h_{2}) \cdot det(1+T^{*}T)^{-2N-1},
\end{equation}

This has consequences for determinants in dimension one, where the
we work with the compact Grassmannian. Let $X=[a_0,a_1 ]$ and let
$E$ be a complex Hermitian bundle over $X$ with unitary connection
$\nabla$. Then the associated generalized Dirac operator is simply
$D = i\nabla_{d/dx}:C^{\infty}(X;E)\to C^{\infty}(X;E).$ Choosing
a trivialization of $E$, so that $E_{a_{0}}\oplus E_{a_{1}} =
\CC^{n}\oplus \CC^{n}$, a global boundary condition $D$ is
specified by an element $P\in Gr (\CC^{n}\oplus \CC^{n})$,
defining the elliptic boundary value problem: $ D_{P} =
i\nabla_{d/dx} :{\rm dom}(D_{P})\too L^{2}([a_0,a_1];E)$.

The Fock functor is here is a topological 0+1-dimensional FQFT
from the category $\Cc_{1}$, whose objects are points endowed with
a complex finite-dimensional Hermitian vector space $V$ (we do not
need to give a polarization in this finite-dimensional situation),
and whose morphisms are compact 1-dimensional manifolds with
boundary with Hermitian bundle with unitary connection. The Fock
functor $Z$ takes an object $(p,V)\in \Cc_{1}$ to the Fock space
$Z(p,V):= \Gamma_{hol}(Gr(V);(\Det(\Ee))^{*})^* \cong \wedge V$,
where $\Ee$ is the usual canonical vector bundle over the
Grassmannian. Consider two compatible morphisms
 $\Lambda_{01} = ([a_0,a_1],E^{01},\nabla^{01})$ and
 $\Lambda_{12} = ([a_1,a_2],E^{12},\nabla^{12})$ in $\Cc_{1}$, so
 that
$$\Lambda_{02} = \Lambda_{12}\Lambda_{01} =
 ([a_0,a_2],E^{02},\nabla^{02}),$$
 with $E^{02}|_{[a_0,a_1]} = E_{01}$ etc. Let $V_{i}$ be the fibre
 over $a_{i}$, and in $[a_i,a_j]$ we assign $a_i$ to be
 `incoming' and $a_j$ to be `outgoing'. For incoming boundary
 components  $a_i$ the associated object in $\Cc_1$ is $(a_i,\ol{V_i})$.
 Then we define  $Z(\Lambda_{ij}) = \nu_{K_{ij}}$,
 where $K_{ij} \in Gr(V_{0}\oplus V_{j})$ is the Calderon
subspace of boundary values of solutions to the `Dirac' operator
$D = i\nabla_{d/dx}^{ij}$. We have $$Z(\Lambda_{ij})\in
Z((p_i,\ol{V}_i)\sqcup (p_j,V_j)) = Z(p_i,\ol{V}_i)\otimes
Z(p_j,V_j) \cong (\wedge \ol{V}_{i})\otimes (\wedge V_{j})\cong
\Hom(\wedge V_{i},\wedge V_{j}) := \Ff_{K_{ij}}.$$

\noi Because $K_{ij} = graph(h_{ij}:V_{i}\to V_{j})$ with $h_{ij}$
 the parallel-transport of the connection on $E_{ij}$ between $a_i$
and $a_j$, a simple computation gives under the above
identification $Z(\Lambda_{ij})\longleftrightarrow \wedge
h_{ij}\in \Hom (\wedge V_{i},\wedge V_{j}).$ Next we have a
canonical pairing
\begin{equation}\label{e:1pairing}
Z(a_0,\ol{V_{0}})\otimes Z(a_1,V_1)\otimes Z(a_1,\ol{V_1})\otimes
Z(a_2,V_2) \too Z(a_0,\ol{V_{0}})\otimes  Z(a_2,V_2),
\end{equation}
\noi induced by subtraction $V_1\oplus V_1 \to V_1$, with
 $\wedge
h_{01} \otimes \wedge h_{12} \too \wedge h_{01}h_{12}.$ If we take
the case where $a_2 = a_0$, so that
 $\Lambda_{02} = \Lambda_{12}\Lambda_{01} = (S^{1} = [a_0,a_2],E^{02},\nabla^{02}),$
 corresponding to morphisms in $Gr(\ol{V_0}\oplus V_1)$ and $Gr(\ol{V_1}\oplus V_0)$
respectively, then $Z(\Lambda_{01})\in\Ff_{K_{01}}$ and
$Z(\Lambda_{10})\in\Ff_{K_{10}^{\perp}},$ and the induced pairing
$\Ff_{K_{01}}\otimes \Ff_{K_{10}^{\perp}} \too \mathbb{C},$ under
the above identifications is just the supertrace $$(\, ,\,): \Hom
(\wedge V_0,\wedge V_1)\otimes \Hom (\wedge V_1,\wedge V_0) \too
\mathbb{C},$$ $$(a,b) \mapsto \tr_{s}(ab) :=
\sum_{k}(-1)^{k}\tr(ab|_{\wedge^{k}}).$$ Applied to the vacuum
elements $\nu_{W_{T_{01}}}\leftrightarrow \wedge T_{01}\in  \Hom
(\wedge V_{0},\wedge V_{1})$ and
$\nu_{W_{T_{10}}^{\perp}}\leftrightarrow \wedge (-T^{*}_{10}) \in
\Hom (\wedge V_{1},\wedge V_{0})$ we have
\begin{equation}\label{e:vacpairing}
  (\nu_{W_{T_{01}}},\nu_{W_{T_{10}}^{\perp}}) = \tr_{s}(\wedge -T^{*}_{10}\wedge
  T_{01}) = \sum_{k}\tr(\wedge^{k}(T^{*}_{10}T_{01}) = det(I+T_{10}^{*}T_{01}).
\end{equation}
\noi Hence we have $ (Z(\Lambda_{01}),Z(\Lambda_{10})) =
det(I-h_{10}h_{01}),$ (since $h_{ij}$ is unitary), and $
(Z(\Lambda_{01}),\nu_{W_{T}^{\perp}}) = det(I+T^{*}h_{01}).$ On
the other hand it well-known that $det(I+T^{*}h_{01}) =
det_{\zeta}(D_{P_{T}}).$ So from eq. \eqref{e:2integral} we have
\begin{equation}\label{e:3integral}
 (Z(\Lambda_{01}),Z(\Lambda_{10})) = a_N \int dT dT^* det_{\z}(D^{10}_{P_{T}})
det_{\z}(D^{01}_{P_{-T^{*}}}) \cdot det(1+T^{*}T)^{-2N-1},
\end{equation}
where $P_{-T^{*}} = I - P_{T}$, expressing the relation of the
algebraic Fock space pairing to the path integral sewing formula
eq. \eqref{e:detsewing}.

Notice that the gauge group of a boundary component of $[a_0,a_1]$
is just a copy of the unitary group $U(n)$ and under the embedding
$g\to graph (g) := W_{g}$, the Fock functor maps $g$ to $\wedge g$
on $\wedge V$. Thus in the case of $0+1$-dimensions the FQFT
representation of the boundary gauge group is the fundamental
$U(n)$-representation, which is a restatement of the Borel-Weil
Theorem for $U(n)$. This means that the `invariant' output by the
FQFT, which in fact here is a TQFT, is the character of the
fundamental representation $\pi$ of $U(n)$. This is what we would
expect. We are dealing with a single particle evolving through
time, and so its only invariants are the representations of its
internal symmetry group, which is the symmetry group of the bundle
$E$ over $[a,b]$. In this sense we are dealing with quantum
mechanics, rather than QFT, and because it is a topological field
theory the Hilbert space is finite-dimensional.

\subsection{Relation to the Berezin integral}
The above pairing can also be described by a Fermionic integral.
Let $\wedge V$ denote the exterior algebra of the complex vector
space $V$ with odd generators $\xi_{1},\ldots,\xi_{n}$. It has
basis the monomials $\xi_{I} =
\xi_{i_{1}}\ldots\xi_{i_{p}},\;\;\;\;\;I=\{i_{1},\ldots,i_{p}\},\;\;\;\;i_{1}<\ldots
< i_{p},$  where $I$ runs over subsets of $\{1,\ldots,n\}$, and we
set $|I|:=p$. The Fermionic (or Berezin) integral is the linear
functional $$\int :\wedge V \too \mathbb{C}, \;\;\;\; f(\xi)
\longmapsto \int f(\xi) \,\Dd\xi $$ \noi which picks out the the
top degree coefficient of $f(\xi)$ (a polynomial in the
generators) relative to the generator $\xi = \xi_{1}\ldots\xi_{n}$
of $\Det V = \wedge^{n}V$. This extends to a functional $$\int
:\wedge \ol{V}\otimes \wedge V \too \mathbb{C}, \;\;\;\;
f(\ol{\xi},\xi) \longmapsto \int f(\ol{\xi},\xi) \,\Dd\xi
\Dd\ol{\xi}, $$ \noi defined relative to the generator
 $ \xi\ol{\xi} := \xi_{1}\ol{\xi_{1}}\ldots\xi_{n}\ol{\xi_{n}}.$
  of $\Det V\otimes \Det\ol{V}$.
Given an element $T\in \End (V)$ we associate to the quadratic
element $\ol{\xi}T\xi := \sum_{i,j}t_{ij}\ol{\xi_{i}}\xi_{j}.$ We
then have $\frac{1}{n!}(\ol{\xi}T\xi)^{n} = det(T)\xi\ol{\xi},$
and more generally the Gaussian expression $$e^{\ol{\xi}T\xi } =
\sum_{I}det(T_{I})\xi_{I}\ol{\xi_{I}},$$ \noi where $T_{I}$
denotes the submatrix $(t_{ij})$ with $i,j\in I$, so that we can
write
\begin{equation}\label{e:FdetA}
  \int e^{\ol{\xi}T\xi } \,\Dd\xi
\Dd\ol{\xi}  = det(T),
\end{equation}
\noi so determinants are expressible as complex Fermion Gaussian
integrals.

Next, we have a bilinear form on $\wedge \ol{V}\otimes \wedge V$
defined by
\begin{equation}\label{e:bform}
  <f,g> = \int g(\ol{\xi},\xi)^{\sigma} f(\ol{\xi},\xi)
  \,\Dd\mu[(\xi,\xi),(\ol{\xi},\ol{\xi})],
\end{equation}
\noi where $f(\xi)^{\sigma}$ is $f(\xi)$ with the order of the
generators reversed, and $$\int f(\ol{\xi},\xi)
\,\Dd\mu[(\xi,\xi),(\ol{\xi},\ol{\xi})] := \int
f(\ol{\xi},\xi)e^{2\ol{\xi}\xi}
  \,\Dd(\xi,\xi)\Dd(\ol{\xi},\ol{\xi}),$$
 \noi is the Fermionic integral with respect to a Gaussian
 measure. The $2$ arises in the exponent because we are dealing
 with $\wedge \ol{V}\otimes \wedge V$ rather than $ \wedge V$.
 Applied to quadratic elements $e^{\ol{\xi}T\xi }$ and  $e^{\ol{\xi}S\xi
 }$ defined for $T,S\in\End(V)$ we then have
 \begin{eqnarray*}
 <e^{\ol{\xi}T\xi },e^{\ol{\xi}S\xi
 }> & = & \int e^{\ol{\xi}S\xi + \ol{\xi}T\xi + 2\ol{\xi}\xi} \,\Dd\xi
\Dd\ol{\xi}\\ & = & \int e^{(\ol{\xi}\;\ol{\xi})
\begin{pmatrix}
  I & T \\
  S & I
\end{pmatrix}\begin{pmatrix}
  \xi \\
  \xi
\end{pmatrix}}
 \,\Dd\xi
\Dd\ol{\xi}\\ & = & det_{V\oplus V}\begin{pmatrix}
  I & T \\
  S & I
\end{pmatrix} \\ & = & det_{V}(I-ST).
\end{eqnarray*}
\noi Here we use \eqref{e:FdetA} applied to $\begin{pmatrix}
  I & T \\
  S & I
\end{pmatrix} :V\oplus V \to V \oplus V$ and the general formula
$det\begin{pmatrix}
  a & b \\
  c & d
\end{pmatrix} = det(d)det(a-bd^{-1}c),$
valid provided $d:V\to V$ is invertible.

We can repeat the process for a pair of complex vector spaces
$V_{0}\neq V_{1}$ of the same dimension and
$T\in\Hom(V_{0},V_{1})$ and $S\in\Hom(V_{1},V_{0}).$ Now define
the Fermionic integral just to be the projection onto the form of
top degree $\int :\wedge \ol{V_{0}}\otimes \wedge V_{1} \to
\Det(V_{0},V_{1}).$ \noi Associated to $T$ we have
$e^{T}\in\ol{V_{0}}\otimes \wedge V_{1}\cong\Hom(\wedge
V_{0},\wedge V_{1})$ we may regard as an element of
$\ol{V_{0}}\otimes \wedge V_{1}$ via the Hermitian isomorphism
$\ol{V_{0}}\cong V_{0}^{*}$, and then $\int e^{T} = det (T)\in
\Det(V_{0},V_{1}).$  Here $det (T)$ is the element
$det(T)(\xi_{1}\ldots\xi_{n})=T\xi_{1}\ldots T\xi_{n}$, for a
basis $\xi_{i}$ of $V_{0}$, which is canonically identified with
$det(T)\in\mathbb{C}$ when $V_{0}=V_{1}$, and the Gaussian element
is then $e^{T} = e^{\ol{\xi}T\xi }$. The bilinear pairing goes
through as before, with  $<e^{T},e^{S}> = det_{V_{0}}(I-ST),$
which gives an alternative formulation of the Fock pairing $<\,
,\,>:\Ff_{W_{T_{0}}}\times \Ff_{W_{T_{1}}^{\perp}}\to \mathbb{C}$.

\end{document}